\documentclass{article}

\usepackage[preprint]{neurips_2023}

\usepackage[utf8]{inputenc} 
\usepackage[T1]{fontenc}    
\usepackage{hyperref}       
\usepackage{url}            
\usepackage{booktabs}       
\usepackage{amsfonts}       
\usepackage{nicefrac}       
\usepackage{microtype}      
\usepackage{xcolor}         
\usepackage{bm}
\usepackage{amsmath}
\usepackage{graphicx}
\usepackage{comment}
\usepackage{listings}
\usepackage{amsthm}
\usepackage{wrapfig}
\usepackage{multirow}

\newtheorem{theorem}{Theorem}

\title{Conditional Perceptual Quality Preserving Image Compression}

\author{
Tongda Xu \\
Air, Tsinghua University \\
\texttt{x.tongda@nyu.edu} \\
\And
Qian Zhang \\
Air, Tsinghua University \\
\texttt{zhangqian@stu.xjtu.edu.cn} \\
\And 
Yanghao Li \\
Air, Tsinghua University \\
\texttt{liyangha18@mails.tsinghua.edu.cn} \\
\And
Dailan He \\
The Chinese University of Hong Kong\\
\texttt{hedailan@sensetime.com} \\
\And
Zhe Wang \\
Air, Tsinghua University \\
\texttt{wangzhe@air.tsinghua.edu.cn} \\
\And
Yuanyuan Wang \\
Sensetime Research \\
\texttt{wangyuanyuan@sensetime.com} \\
\And
Hongwei Qin \\
Sensetime Research \\
\texttt{qinhongwei@@sensetime.com} \\
\And
Yan Wang \\
Air, Tsinghua University \\
\texttt{wangyan@air.tsinghua.edu.cn} \\
\And
Jingjing Liu \\
Air, Tsinghua University \\
\texttt{JJLiu@air.tsinghua.edu.cn} \\
\And
Ya-Qin Zhang \\
Air, Tsinghua University \\
\texttt{zhangyaqin@air.tsinghua.edu.cn} \\
}

\begin{document}
 
\maketitle

\begin{abstract}
We propose conditional perceptual quality, an extension of the perceptual quality defined in \citet{blau2018perception}, by conditioning it on user defined information. Specifically, we extend the original perceptual quality $d(p_{X},p_{\hat{X}})$ to the conditional perceptual quality $d(p_{X|Y},p_{\hat{X}|Y})$, where $X$ is the original image, $\hat{X}$ is the reconstructed, $Y$ is side information defined by user and $d(.,.)$ is divergence.
We show that conditional perceptual quality has similar theoretical properties as rate-distortion-perception trade-off \citep{blau2019rethinking}. Based on these theoretical results, we propose an optimal framework for conditional perceptual quality preserving compression. Experimental results show that our codec successfully maintains high perceptual quality and semantic quality at all bitrate. Besides, by providing a lowerbound of common randomness required, we settle the previous arguments on whether randomness should be incorporated into generator for (conditional) perceptual quality compression. The source code is provided in supplementary material.
\end{abstract}

\section{Introduction}

How to optimize image compression algorithms to preserve perceptual quality is a fundamental problem. As mean square error (MSE) leads to blurry results, various loss functions are designed including SSIM \citep{wang2004image} and LPIPS \citep{zhang2018unreasonable}. In 2018, \citet{blau2018perception} formalize perceptual quality as the divergence of original image distribution and distorted image distribution. Specifically, given an original image $X$ and distorted image $\hat{X}$, they evaluate perceptual quality as $d(p_{X},p_{\hat{X}})$, where $d(.,.)$ is the divergence. Later, their theory is developed into rate-distortion-perception trade-off \citep{blau2019rethinking}. Since then, rate-distortion-perception trade-off has become the theoretical cornerstone for perceptual quality preserving image compression \citep{agustsson2019generative,Mentzer2020HighFidelityGI,Agustsson2022MultiRealismIC}, which also provides a theoretical justification of previous codecs using generative adversarial network \citep{Rippel2017RealTimeAI,tschannen2018deep}.

However, perceptual quality preserving compression alone is not enough. Consider we have a codec for MNIST dataset \citep{LeCun2005TheMD}. At low bitrate, a perceptual quality preserving codec might decode an image of "7" from an original image of "3" (See Fig.~\ref{fig:mnisteg}). Though the decoded image of "7" has perfect perceptual quality, the digit is incorrect. Another example is in \citet{tschannen2018deep}, where a perceptual quality preserving codec might compress the image of one bedroom into another with completely different layout. Though the decoded bedroom has perfect perceptual quality, the layout of bedroom is wrongly depicted.

In both cases, there is specific side information that we want to keep in the original images. Let's denote this information as $Y$. For the MNIST case, $Y$ is the digit. And for the bedroom case, $Y$ is the layout. In both bases, the perceptual quality is perfect, and the divergence between original image and distorted image $d(p_{X},p_{\hat{X}})=0$. However, the divergence between the images' posterior on $Y$ is obviously different, indicating a large divergence $d(p_{X|Y},p_{\hat{X}|Y})$ between conditional distributions.

In this paper, we propose conditional perceptual quality, which considers the divergence between conditional distributions $d(p_{X|Y},p_{\hat{X}|Y})$ instead of $d(p_{X},p_{\hat{X}})$. We show that such conditional extension shares similar desirable theoretical properties to rate-distortion-perception trade-off \citep{blau2019rethinking}. Based on those theoretical properties, we propose an optimal coding framework to achieve perfect conditional perceptual quality. We empirically show that our coding framework can achieve high perceptual quality and semantic quality at all bitrate with bounded distortion. And it smoothly interpolates the high-to-low distortion as bitrate increases. Besides, we settle the argument on whether noise should be incorporated into decoder for perceptual quality preserving compression, by proving a lowerbound of noise dimensions required for different rate.

\section{Preliminaries}
\subsection{Rate Distortion Theory}

Consider the concatenation of $n$ i.i.d. source images $X^{n}=\{X_1,...,X_n\}$ and an encoder $f^{n}(.)$, we encode $X^n$ into an index $ M^n\in\{1,...,2^{nR^{n}}\}$ with $nR^n$ bits. Specifically, $R^n = \log|\hat{\mathcal{X}}^n|/n$, and $|\hat{\mathcal{X}}^n|$ is the size of reconstruction alphabet. With a decoder $g^n(.)$, we can obtain the reconstruction image $\hat{X}^n = g^n(M^n) \in \hat{\mathcal{X}}^n$. As in \citet{cover1999elements}, we define information rate distortion function as 
\begin{align}
    R^{I}(D) = \min_{p_{\hat{X}|X}} I(X,\hat{X}), s.t. \mathbb{E}[\Delta(X,\hat{X})] \le D \label{eq:rdi},
\end{align}
where $\Delta(.,.)$ is distortion and $D$ is the distortion constraint. Rate distortion theorem \citep{cover1999elements} tells us that when $n\rightarrow\infty$, there exists an encoder-decoder (codec) $f^n(.),g^n(.)$ with $\mathbb{E}[\Delta(X^n,\hat{X}^n)] \le D$ if $R^{n}> R^{I}(D)$, and no codec with $R^{n}< R^{I}(D)$.

Now, consider a single source image $X \in \mathcal{X}$ and an encoder $f^{0}(.)$, we encode $X$ into uniquely decodable code $M^0=f^{0}(X)$. With a decoder $g^{0}(.)$, we can obtain a reconstruction image $\hat{X} = g^{0}(M^0)$. We denote the expected code length of $M^0$ as $R^{0} = \mathbb{E}[\mathcal{L}(M^0)]$. As in \citet{li2018strong}, we say $(R^0,D)$ is one-shot achievable if there exists a one-shot codec $f^0(.),g^0(.)$ with rate $R^{0}$, distortion $\mathbb{E}[\Delta(X,\hat{X})] \le D$. And \citet{li2018strong} show that any $(R^0,D)$ with
\begin{align}
    R^0 > R^I(D) + \log (R^I(D) + 1) + 6
\end{align}
is achievable.
\subsection{Rate-Distortion-Perception Trade-off}
\citet{blau2018perception} formalize a metric for perceptual quality as the divergence between two image distributions $d(p_X,p_{\hat{X}})$, where $p_X$ is the distribution for the original image, $p_{\hat{X}}$ is the distribution of reconstruction image and $d(.,.)$ is the divergence. They show that $d(p_X,p_{\hat{X}})$ is at odds with any distortion $\Delta(X,\hat{X})$ between two specific images. Since then, this definition of perceptual quality has been widely applied to image restoration, super-resolution and compression. Later, \citet{blau2019rethinking} extend the information rate distortion function $R^I(D)$ into rate-distortion-perception function as:
\begin{align}
    R^{I}(D,P) = \min_{p_{\hat{X}|X}} I(X,\hat{X}), s.t. \mathbb{E}[\Delta(X,\hat{X})] \le D,d(p_{X},p_{\hat{X}})\le P,
\end{align}
where $P$ is the constraint on divergence. The fundamental properties of $R^{I}(D,P)$, such as monotonicity, convexity, achievability and converse, and one-shot achievability, are proven by \citet{blau2019rethinking,theis2021coding,Zhang2021UniversalRR} in analogous to rate distortion. Beyond them, more useful properties are also shown:
\begin{theorem}
\label{thm:1}
When distortion $\Delta(.,.)$ is MSE:
\begin{enumerate}
    \item \citep{blau2019rethinking} $R^{I}(D,0) \le R^{I}(\frac{1}{2}D,\infty) = R^{I}(\frac{1}{2}D)$;
    \item \citep{Yan2021OnPL} Given an optimal one-shot codec $f^0(.),g^0_1(.)$ with code $M^0$, reconstruction $\hat{X}$ and distortion $\mathbb{E}[\Delta(X,\hat{X})]\le D/2$, there exists an optimal perceptual decoder $g_2(.)$, with $ p_{\hat{X}|M^0}= p_{X|M^0},d(p_X,p_{\tilde{X}})=0,\mathbb{E}[\Delta(X,\tilde{X})]\le D$;
\end{enumerate}
\end{theorem}
Prior to Theorem. 1, we only know that achieving perfect perceptual quality increases MSE \citep{tschannen2018deep}, and there are arguments on whether we should enable the gradient of encoder while training perceptual decoder \citep{Rippel2017RealTimeAI,Mentzer2020HighFidelityGI}. Theorem.~\ref{thm:1}.1 and Theorem.~\ref{thm:1}.2 show that we can achieve perfect conditional perceptual quality by first training an MSE codec and then a perceptual decoder, and the cost is at most doubling MSE.

\section{Conditional Perceptual Quality Preserving Image Compression}
\begin{figure}[t]
\centering
 \includegraphics[width=1.0\linewidth]{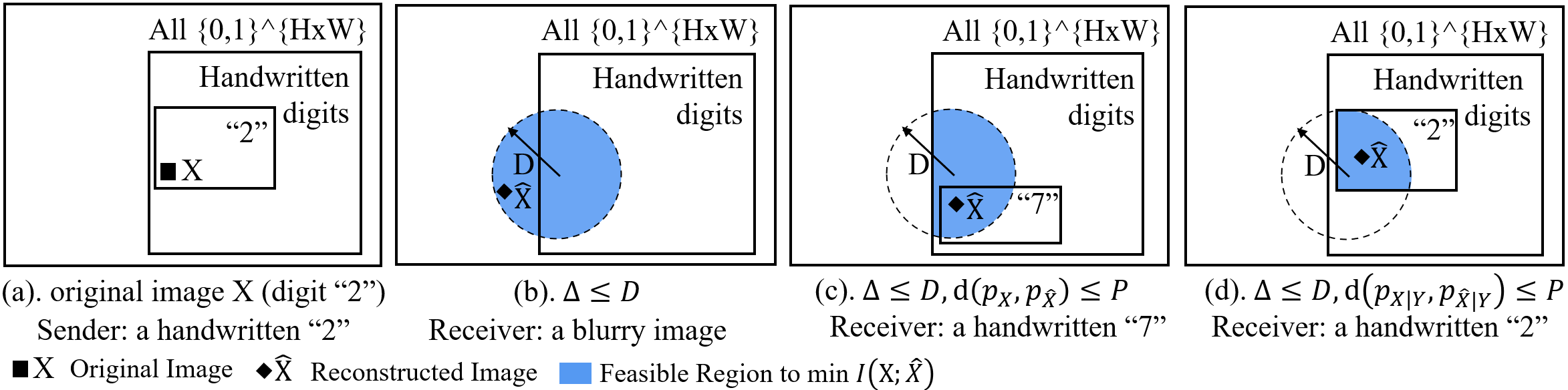}
 \caption{Feasible region of: (a). original image; (b). $R^I(D)$; (c). $R^I(D,P)$; (d). $R^I_C(D,P)$.}
 \label{fig:eg}
\end{figure}
\subsection{Rate-Distortion-Conditional Perception Trade-off} We extend the perceptual quality defined by \citet{blau2018perception} by considering the divergence between the posteriors of images $d(p_{X|Y},p_{\hat{X}|Y})$, where $Y$ is the user-defined side information. We name it `\textit{conditional perceptual quality}'. For now we assume $Y$ is shared across encoder and decoder and ignore its bitrate. Later in Theorem.~\ref{thm:04}, we will show the overhead of this assumption is no more than $2$ bits. Similar to rate-distortion-perception function, we define rate-distortion-conditional perception function as:
\begin{align}
    R^{I}_C(D,P) = \min_{p_{\hat{X}|X,Y}} I(X;\hat{X}|Y), s.t. \mathbb{E}[\Delta(X,\hat{X})|Y] \le D,d(p_{X|Y},p_{\hat{X}|Y})\le P.
\end{align}
To better understand the relationship between $R^I(D),R^I(D,P),R^I_C(D,P)$ and why the conditional perceptual quality makes sense, we revisit the MNIST example. Consider the sender has an image of digit "2" $X$ to send, and all $R^I(D),R^I(D,P),R^I_C(D,P)$ minimize rate $I(X;\hat{X})$ within their feasible region. For $R^I(D)$, the feasible region is a sphere (Fig.~\ref{fig:eg}.(b)) and not all images within it are handwritten digits. It is likely that $\hat{X}$ is a blurry image. On the other hand, constraining perceptual quality $d(p_{X},p_{\hat{X}})\le P$ makes the feasible region an intersection of all handwritten digits and the sphere (Fig.~\ref{fig:eg}.(c)). For $R^I(D,P)$, all possible $\hat{X}$s are handwritten images, but $\hat{X}$ might have different digits. Now consider constraining the conditional perceptual quality $d(p_{X|Y},p_{\hat{X}|Y})\le P$, then the feasible region becomes an intersection of all handwritten "2" and sphere (Fig.~\ref{fig:eg}.(d)). Then for $R^I_C(D,P)$, all possible $\hat{X}$ can be recognized as a handwritten "2" by the receiver.

Now we show that $R^{I}_C(D,P)$ shares similar theoretical property as $R^I(D,P)$:
\begin{theorem}
\label{thm:2}
$R^{I}_C(D,P)$ shares similar fundamental property to $R^I(D),R^I(D,P)$:
\begin{enumerate}
    \item (monotonicity and convexity) $R^I_C(D,P)$ is monotonously non-increasing in D,P. And when $d(.,.)$ is convex in its second argument, $R^I_C(D,P)$ is convex.
    \item (one-shot achievability) There exists a one-shot codec $f^0(.,Y),g^0(.,Y)$ that satisfies $\mathbb{E}[\Delta(X,\hat{X})|Y]\le D,d(p_{X|Y},p_{\hat{X}|Y})\le P$ if $R^{0}> R^I_C(D,P) + \log (R^I_C(D,P) + 1) + 5$;
    \item (achievability and converse) When $n\rightarrow\infty$, there exists a codec $f^n(.,Y),g^n(.,Y)$ that satisfies $\mathbb{E}[\Delta(X^n,\hat{X}^n)|Y]\le D,d(p_{X|Y},p_{\hat{X}|Y})\le P$ if  $R^{n}> R^{I}_C(D,P)$ and no codec with $R^{n}<R^{I}_C(D,P)$;
    \item Perfect conditional perceptual quality leads to perfect perceptual quality. 
\end{enumerate}
Furthermore, when distortion $\Delta(.,.)$ is MSE:
\begin{enumerate}
\setcounter{enumi}{4}
    \item  $R^{I}_C(D,0) \le R^{I}_C(\frac{1}{2}D,\infty)$;
    \item Given an optimal one-shot codec $f^0(.,Y),g^0_1(.,Y)$ with code $M^0$, reconstruction $\hat{X}$ and distortion $\mathbb{E}[\Delta(X,\hat{X})|Y]\le D/2$, there exists an optimal perceptual decoder $g_2(.,Y)$ with $ p_{\tilde{X}|M^0,Y}= p_{X|M^0,Y}, d(p_{X|Y},p_{\tilde{X}|Y})=0,\mathbb{E}[\Delta(X,\tilde{X})|Y]\le D$;
\end{enumerate}
\end{theorem}

Theorem.~\ref{thm:2} provides bound and guideline for conditional perceptual quality preserving codec. More specifically, Theorem.~\ref{thm:2}.5, Theorem.~\ref{thm:2}.6 show that we can achieve perfect conditional perceptual quality by training first an MSE codec and then a conditional perceptual decoder, and the cost is at most doubling MSE.

\subsection{Conditional Perceptual Quality Preserving Image Compression}

In previous section, we assume $Y$ is shared across encoder and decoder, and we define $R^I_C(D,P)$ without $Y$'s bitrate. Now as we are designing practical zero-shot codec, we take the bitrate of $Y$ into consideration. Consider the following coding framework:
\begin{enumerate}
    \item We encode $Y$ losslessly with expected code length $R_Y$ and share it with decoder;
    \item We generate the code $M=f(X,Y)$ and encode $M$ under the condition of $Y$ with expected code length $R_M$;
    \item The decoder first decodes $Y$ and then with $Y$ it decodes $M$, and with $M$ it decodes $\hat{X}=g(M,Y)$. 
\end{enumerate}
The total expected code length is $R_Y+R_M$. We show that under some mild assumptions, the above coding framework is at most $2$ bits from optimal.
\begin{theorem}
\label{thm:04} Assume $d(p_{X|Y},p_{\hat{X}|Y})=0$ and $Y$ is deterministic of $X$, for any codec with total expected code length $R$ and code $M$, there exists another codec that has the same code $M$ following the above framework with $Y$ encoded losslessly, whose total expected code length $R_M+R_Y \le R+2$.
\end{theorem}
In practice, even for MSE optimized codec with $d(p_{X|Y},p_{\hat{X}|Y})$ constraint $P=\infty$, the performance decay of the above coding framework is small, as the net bitrate of $Y$ is usually small. By far, we have shown that the above coding scheme is at most $2$ bits from optimal (Theorem.~\ref{thm:04}), and the bitstream of MSE codec and conditional perceptual codec can be shared (Theorem.~\ref{thm:2}.5). Then, we can extend the perceptual quality preserving codec \citep{Yan2021OnPL} to conditional perceptual quality preserving codec, by encoding $Y$ losslessly and adding a conditional entropy model for code $M$.

\begin{figure}[t]
\centering
 \includegraphics[width=0.8\linewidth]{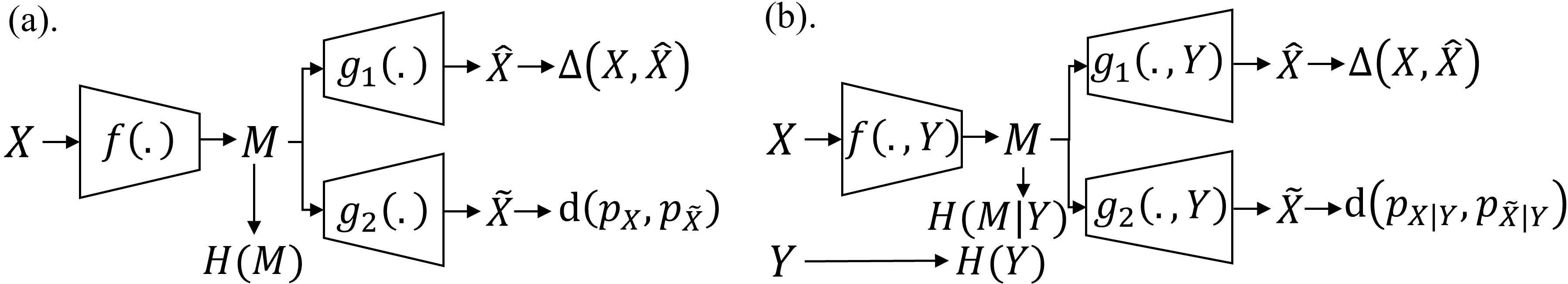}
 \caption{(a). The perceptual quality preserving framework by \citet{Yan2021OnPL}. (b). The conditional perceptual quality preserving framework by us.}
 \label{fig:fw}
\end{figure}

Specifically, for perceptual quality preserving codec (Fig.~\ref{fig:fw}.(a)), \citet{Yan2021OnPL} propose the following optimization framework:
\begin{enumerate}
    \item They first optimize an MSE codec $f(.), g_1(.)$ using rate distortion loss $\mathbb{E}[H(M)+\lambda \Delta(X,\hat{X})]$, where $M=f(X),\hat{X}=g_1(M)$, $\lambda$ is the Lagrangian multiplier controlling rate-distortion trade-off, and $H(M)$ is the entropy of code $M$;
    \item Then they optimize a perceptual quality preserving decoder $g_2(.)$ with $d(p_{X|M},p_{\tilde{X}|M})=0$ constraint implemented by a conditional generative model.
\end{enumerate}
Theorem.~\ref{thm:1}.2 shows that ideally this leads to $d(p_{X},p_{\tilde{X}})=0$ and $\mathbb{E}[\Delta(X,\tilde{X})]\le 2\mathbb{E}[\Delta(X,\hat{X})]$.

Similarly, for conditional perceptual quality preserving codec (Fig.~\ref{fig:fw}.(b)), we combine the proposed optimal conditional perceptual coding framework with the perceptual quality preserving optimization framework, and propose the following optimization framework:
\begin{enumerate}
    \item We encode $Y$ losslessly and share it with the decoder with rate close to $H(Y)$.
    \item We first optimize an MSE codec $f(.,Y), g_1(.,Y)$ using rate distortion loss $\mathbb{E}[H(M|Y)+\lambda \Delta(X,\hat{X})]$. 
    \item We then optimize a conditional perceptual quality preserving decoder $g_2(.,Y)$, with $d(p_{X|M,Y},p_{\tilde{X}|M,Y})=0$ constraint implemented by a conditional generative model.
\end{enumerate}
Theorem.~\ref{thm:2}.5 shows that ideally this leads to $d(p_{X|Y},p_{\tilde{X}|Y})=0$ and $\mathbb{E}[\Delta(X,\tilde{X})|Y]\le 2\mathbb{E}[\Delta(X,\hat{X})|Y]$. And Theorem.~\ref{thm:04} guarantees that the resulting codec is at most $2$ bits from optimal.
\subsection{Common Randomness Lowerbound} 
Finally, one practical issue that has not been well understood is the random noise $W$ injected at $g_2(.)$ for perfect (conditional) perceptual quality. Previous works in perceptual compression empirically show $W$ is required for low bitrate \citep{tschannen2018deep,Yan2021OnPL}, not required for high bitrate \citep{tschannen2018deep, Mentzer2020HighFidelityGI,Agustsson2022MultiRealismIC}, and is beneficial in a tractable example \citet{theis2021advantages}. To settle those arguments, we quantify the least amount of randomness required to achieve perfect (conditional) perceptual quality for one-shot codec in the next theorem.
\begin{theorem}
    \label{thm:05}
    For one-shot codec, to achieve perfect perceptual quality at rate $R^0$, we require a common randomness $W \in \mathcal{W}$ with at least $\log |\mathcal{W}| \ge H(W) \ge H(X) - (R^0 + 1)$. To achieve perfect conditional perceptual quality, we require $\log |\mathcal{W}| \ge H(W) \ge H(X|Y) - (R^0 + 1)$.
\end{theorem}
Intuitively, Theorem.~\ref{thm:05} tells us that the amount of bitrate reduced by lossy codec is the amount of common randomness required. And this result is aligned with previous empirical results. When bitrate $R^0$ is high, the required $|\mathcal{W}|$ is small and sometimes can be ignored. When $R^0$ is low, the required $|\mathcal{W}|$ can be too large to be ignored, and thus becomes necessary.

\section{Experiments}
\subsection{Experiment Setup}
For all the experiments, we assume $Y$ is deterministic of $X$, and our target is perfect conditional perceptual quality, which means $d(p_{X|Y},p_{\hat{X}|Y})=0$. All the experiments are conducted on a computer with AMD EPYC 7742 64-Core Processor and 8 Nivida A30 GPU. All the code is implemented with Python 3.10 and Pytorch 1.13. The dataset, data type of $Y$, baselines, details of encoder-decoder architecture are specified in each subsections.

Ideally, we evaluate the total bitrate $R$, distortion $\mathbb{E}[\Delta(X,\hat{X})|Y]$ and divergence $d(p_{X|Y},p_{\hat{X}|Y})$. The bitrate is evaluated by entropy and implemented via arithmetic coding \citep{rissanen1979arithmetic}. The distortion is MSE. However, there is no direct way to evaluate $d(p_{X|Y},p_{\hat{X}|Y})$. The common method to approximate the perceptual quality $d(p_{X},p_{\hat{X}})$ is Fr\'echet Inception Distance (FID) $d_{F}(p_{X},p_{\hat{X}})$. Similarly, for $Y$ that is tractable to enumerate, we enumerate each $y_i\in Y$, evaluate prior $p_{Y=y}$, FID $d_F(p_{X|Y=y},p_{\hat{X}|Y=y})$, and $d_{F}(p_{X|Y},p_{\hat{X}|Y})=\sum p_{Y=y}d_F(p_{X|Y=y},p_{\hat{X}|Y=y})$ as an approximation to $d(p_{X|Y},p_{\hat{X}|Y})$. We name this metric as conditional FID (ConFID). In addition, we also use FID to evaluate the perceptual quality $d_(p_{X},p_{\hat{X}})$, and a pre-trained model to predict $Y$ from $\hat{X}$ to evaluate the accuracy $d(p_{Y|X},p_{Y|\hat{X}})$. As when both of them are $0$, $d(p_{X|Y},p_{\hat{X}|Y})=0$. Thus, combining those two might give us a hint about $d(p_{X|Y},p_{\hat{X}|Y})$.

\subsection{Evaluation on MNIST Dataset}
\textbf{Setup} We first evaluate our method on MNIST dataset \citep{LeCun2005TheMD} with $Y$ as the digit. This is the MNIST example we mentioned previously, where the users want perfect perceptual quality with correct digits and minimal rate. The baseline MSE codec and the training details are exactly the same as \citet{Yan2021OnPL}. The difference is: we train our MSE codec $g_1$ with $Y$ available to decoder and entropy model of $M$; we train our perceptual decoder $g_2$ with $Y$ available to decoder and discriminator. And $Y$ is losslessly encoded within $\log 10$ bits. We evaluate the rate, MSE, ConFID, FID and classification accuracy. The digits predicted for accuracy is from a model pre-trained on training dataset (See Appendix.~\ref{app:eset} for details). 
\begin{figure}[thb]
\centering
 \includegraphics[width=0.8\linewidth]{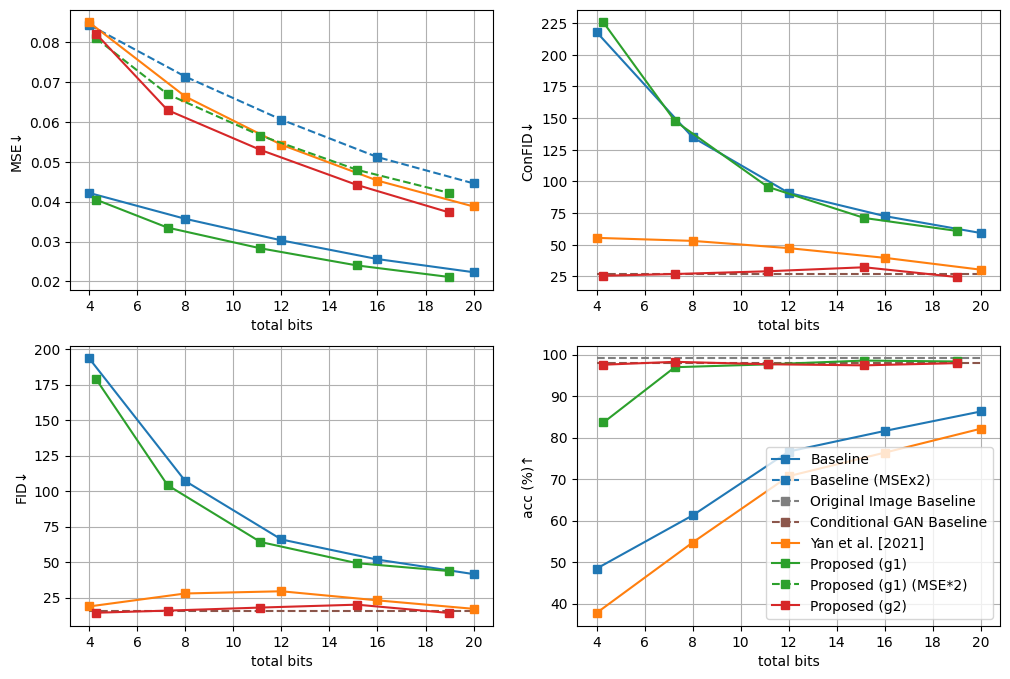}
 \caption{The bitrate-MSE, bitrate-ConFID, bitrate-FID, bitrate-accuracy of different methods. }
 \label{fig:mni}
\end{figure}

\begin{table}[htb]
\centering
\caption{The BD-MSE, BD-ConFID, BD-FID, BD-acc of different methods on MNIST dataset.}
 \label{tab:mni}
\resizebox{\linewidth}{!}{
\begin{tabular}{@{}llllll@{}}
\toprule
 Methods & Constrains & BD-MSE $\downarrow$ &  BD-ConFID $\downarrow$ & BD-FID $\downarrow$ & BD-acc $\uparrow$ \\ \midrule
 Baseline & $\mathbb{E}[\Delta]\le D$ & 0.0000 & 0.00 & 0.00 & 0.00 \\
 \citet{Yan2021OnPL} & $\mathbb{E}[\Delta]\le D, p_X=p_{\hat{X}}$ & 0.0229  & -48.99 & -47.67 & -6.81 \\
 Proposed ($g_1$) & $\mathbb{E}[\Delta]\le D$ & \textbf{-0.0009} & 9.65 & 3.21 & 29.22 \\
 Proposed ($g_2$) & $\mathbb{E}[\Delta]\le D, p_{X|Y}=p_{\hat{X}|Y}$ & 0.0227 & \textbf{-62.15} & \textbf{-50.75} & \textbf{31.19} \\ \bottomrule
\end{tabular}
}
\end{table}

\begin{figure}[thb]
\centering
 \includegraphics[width=0.8\linewidth]{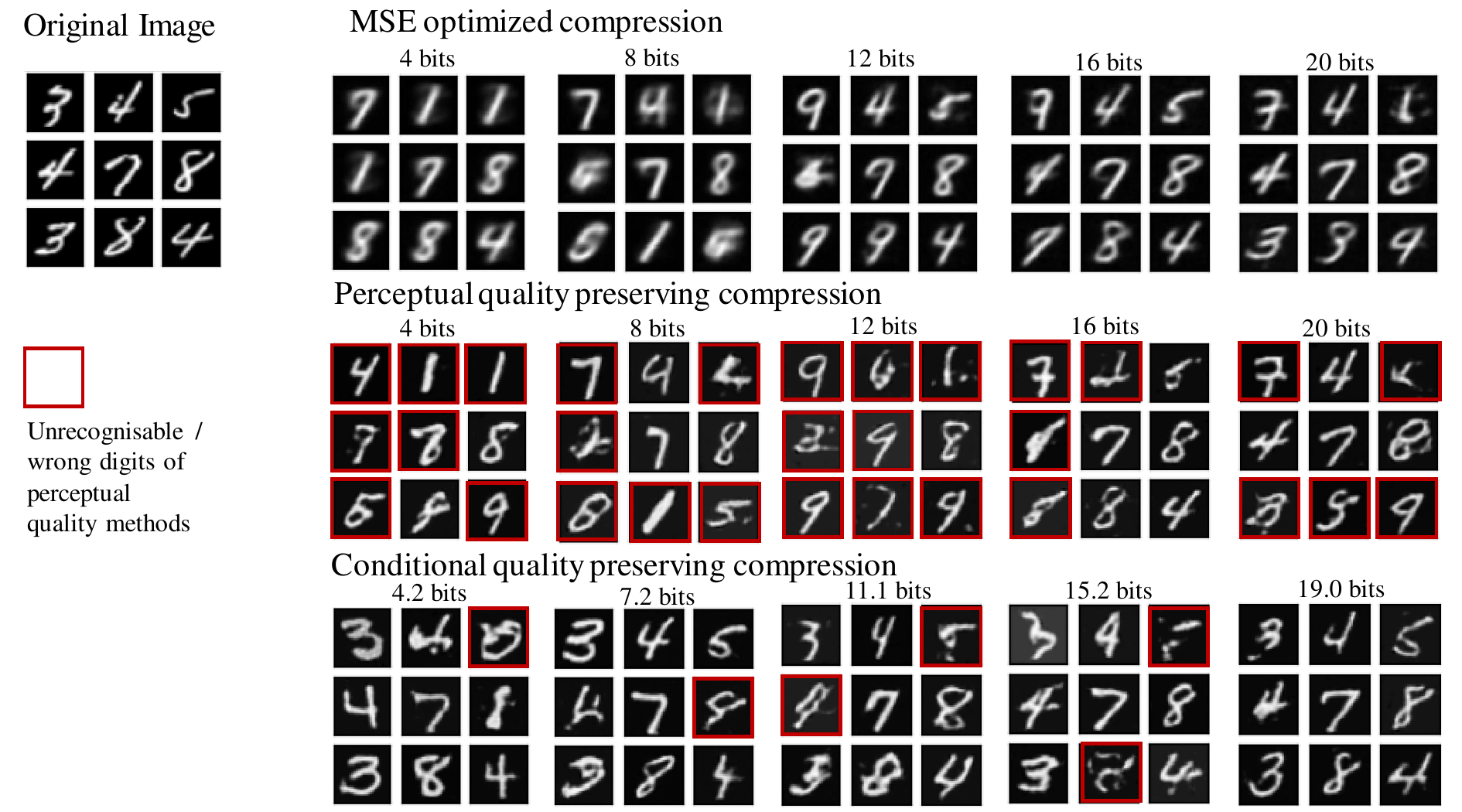}
 \caption{The qualitative results of different methods on MNIST dataset. We mark obvious unrecognisable / wrong digits of two perceptual quality preserving methods in red.}
 \label{fig:mnisteg}
\end{figure}
\textbf{Quantitative Results} We show the rate, distortion in MSE, ConFID, FID and classification accuracy of different methods in Fig.~\ref{fig:mni} and Tab.~\ref{tab:mni}. It is shown that our framework optimized for MSE (Proposed ($g_1$)) is on par with the baseline approach in terms of BD-MSE (-0.0009\% vs 0.0000\%). Surprisingly, our framework optimized for conditional perceptual quality (Proposed ($g_2$)) slightly outperforms \citet{Yan2021OnPL} in terms of BD-MSE and BD-FID (0.0227\% vs 0.0229\%, -50.75\% vs -47.67\%). This means that Proposed ($g_2$) achieves good rate-distortion-perception trade-off. This unexpected result might due to the fact that the digit of an image is an important feature that is more efficient to store than the feature learned by auto-encoder. On the other hand, Proposed ($g_2$) has significant advantage over all other methods in terms of BD-ConFID (-62.15\% vs 0.00\%,-48.99\%,9.65\%), and significantly outperforms baseline and \citet{Yan2021OnPL} in terms of BD-acc (31.19\% vs 0.00\%,-6.81\%). Those results clearly show the advantage of the Proposed ($g_2$) in terms of rate-distortion-conditional perceptual quality trade-off. In addition, Proposed ($g_2$)'s MSE lies below the double of Proposed ($g_1$)'s MSE, which conforms the theoretical results in Theorem.~\ref{thm:2}.

\textbf{Qualitative Results} We show the qualitative comparison of different methods in Fig.~\ref{fig:mnisteg}. It can be seen that the MSE-optimized image is blurry and un-recognisable at low bitrate. On the other hand, the perceptual preserving \citep{Yan2021OnPL} image is visually pleasing, while it often leads to wrong digits. And our proposed conditional perceptual preserving codec (Proposed $g_2$) leads to visually pleasing image with more accurate digit at the same time.

\begin{wrapfigure}{r}{0.4\linewidth}
\centering
\vspace{-1.5em}
 \includegraphics[width=\linewidth]{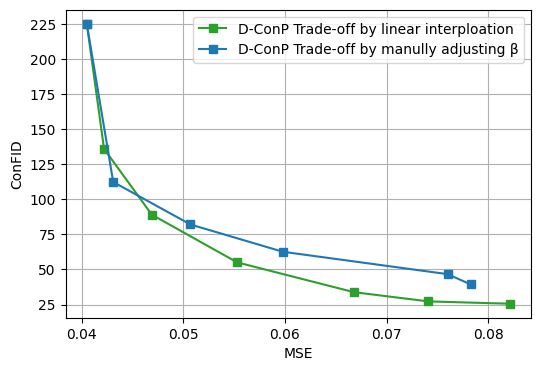}
 \vspace{-2.0em}
 \caption{The MSE-ConFID trade-off.}
 \vspace{-1.0em}
 \label{fig:dpt}
\end{wrapfigure}
\textbf{Distortion-Conditional Perception Trade-off}
Though Theorem.~\ref{thm:2}.6 show that it is possible to achieve perfect (conditional) perceptual quality with MSE encoder, it is unclear whether other middle points between $P=0$ and $P=\infty$ can be achieved with the same bitstream. Existing theoretical results \citep{Zhang2021UniversalRR,Freirich2021ATO,Yan2022OptimallyCP} assume $d(.,.)$ to be Wassertein-2 (W2) distance. This is not very useful for practical codec as W2 does not have a GAN form and requires special tricks to optimize \citep{korotin2019wasserstein}. 

However, it remains alluring to test the method proposed in \citep{Freirich2021ATO,Yan2022OptimallyCP} empirically. More specifically, \citet{Freirich2021ATO,Yan2022OptimallyCP} show that when $d(.,.)$ is W2 distance, any perception-distortion trade-off can be achieved by convex combination between MSE optimized image $\hat{X}$ and perfect perceptual quality image $\tilde{X}$. This requires only $2$ decoder and saves users from complicated decoder manipulation \citep{Iwai2020FidelityControllableEI,Agustsson2022MultiRealismIC}. We linearly interpolate the image reconstructed from $g_1$ and $g_2$, and show their MSE-ConFID. The baseline approach is to jointly train the encoder and decoder to minimize $R+\lambda \mathbb{E}[\Delta(X,\hat{X}))]+\beta d(p_{X|Y},p_{\hat{X}|Y})$, with D-P trade-off achieved by adjusting $\beta$. Fig.~\ref{fig:dpt} shows that simple linear interpolation of images outperforms direct joint training. All the $(D,P)$ points have very similar bitrate and the details are listed in Appendix.~\ref{app:quant}.

\textbf{Evaluation on Common Randomness Required} In additional to the proposed codec, we also verify the correctness of Theorem 4, which is about the lowerbound of common randomness for perfect (conditional) perceptual quality. The details are provided in Appendix.~\ref{app:evcr}.

\subsection{Evaluation on Cityscape Dataset}
\textbf{Setup} We evaluate our method on Cityscape dataset \citep{Cordts2016TheCD} with $Y$ as the  segmentation map down-sampled by $8$. This is similar to the bedroom layout example we mention in previous section, where the users want perfect perceptual quality with roughly correct layout. The baseline MSE codec is \citet{balle2018variational}. For GAN, we adopt the generator and discriminator structure of \citet{Sushko2020YouON}. For perceptual codec \citet{Yan2021OnPL}, we train GAN unconditionally. For conditional perceptual codec, we train GAN conditioned on $Y$, which is compressed losslessly by BMF \citep{bmfref} into $0.00742$ bpp. Different from MNIST, $Y$ is not enumerable and ConFID is not available. Thus, we evaluate the rate, MSE, FID and segmentation mIoU. The segmentation map predicted for mIoU is computed by a model \citep{Yu2017} pre-trained on Cityscape training dataset (See Appendix.~\ref{app:eset} for details). 
\begin{figure}[thb]
\centering
 \includegraphics[width=0.8\linewidth]{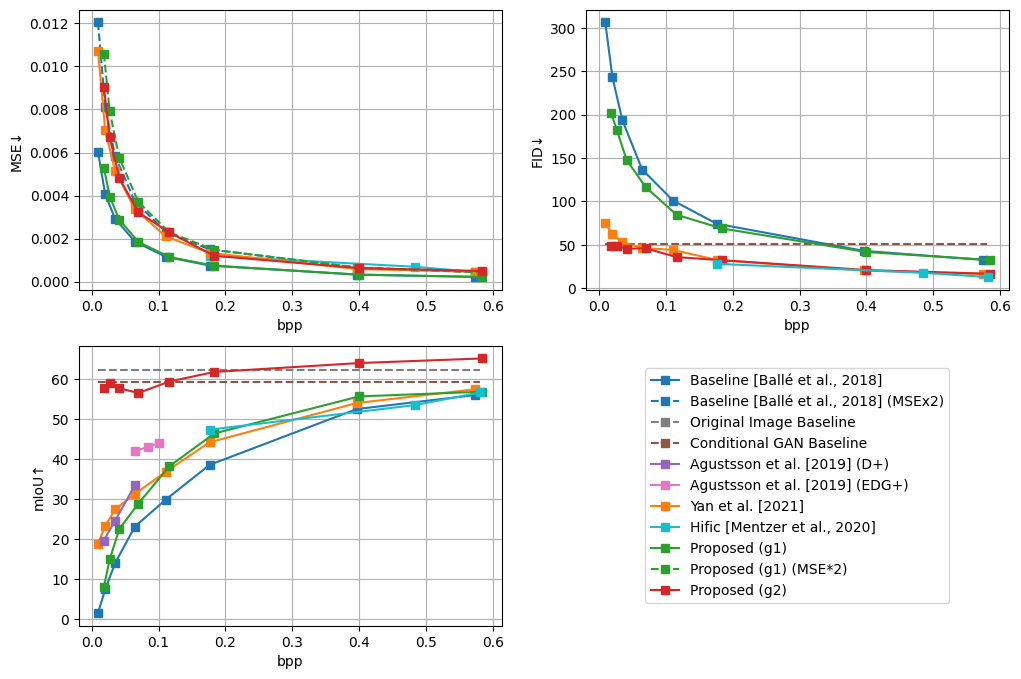}
 \caption{The bitrate-MSE, bitrate-FID, bitrate-mIoU of different methods.}
 \label{fig:csp}
\end{figure}
\begin{table}[htb]
\centering
 \caption{The BD-MSE, BD-FID, BD-mIoU of different methods on Cityscape dataset.}
 \label{tab:csp}
 \resizebox{\linewidth}{!}{
\begin{tabular}{@{}lllll@{}}
\toprule
 Methods & Constrains & BD-MSE $\downarrow$  & BD-FID $\downarrow$ & BD-mIoU $\uparrow$ \\ \midrule
 Baseline \citep{balle2018variational} & $\mathbb{E}[\Delta]\le D$ & 0.0000 & 0.00 & 0.00 \\
 \citet{agustsson2019generative} (D+) & complex & - & - & 11.01 \\
  \citet{agustsson2019generative} (EDG+) & complex & - & - & 16.89 \\
 Hific \citet{Mentzer2020HighFidelityGI}  & complex & 0.0022 & -132.28 & 9.87 \\
 \citet{Yan2021OnPL}  & $\mathbb{E}[\Delta]\le D, p_X=p_{\hat{X}}$ & 0.0022 & -132.28 & 9.87 \\
 Proposed ($g_1$) & $\mathbb{E}[\Delta]\le D$ & \textbf{-0.0004} & -62.64 & 4.97 \\
 Proposed ($g_2$) & $\mathbb{E}[\Delta]\le D, p_{X|Y}=p_{\hat{X}|Y}$ & 0.0017 & \textbf{-167.24} & \textbf{40.32} \\ \bottomrule
\end{tabular}
}
\end{table}

\textbf{Quantitative Results} We show the rate, distortion in MSE, FID and segmentation mIoU of different methods in Fig.~\ref{fig:csp} and Tab.~\ref{tab:csp}. Our framework optimized for MSE (Proposed ($g_1$)) is on par with the baseline approach in terms of BD-MSE (-0.0004\% vs 0.0000\%). And again, our framework optimized for conditional perceptual quality (Proposed ($g_2$)) marginally outperforms \citet{Yan2021OnPL} in terms of BD-MSE and BD-FID (0.0017\% vs 0.0022\%, -167.24\% vs --132.28\%). On the other hand, Proposed ($g_2$) has significant advantage over all other methods in terms of BD-mIoU (40.32\% vs 0.00\%,11.01\%,,16.89\%,9.87\%,4.97\%). In addition, Proposed ($g_2$)'s MSE lies below the double of Proposed ($g_1$)'s MSE, which conforms the the theoretical results in Theorem.~\ref{thm:2}.

\begin{figure}[thb]
\centering
 \includegraphics[width=0.9\linewidth]{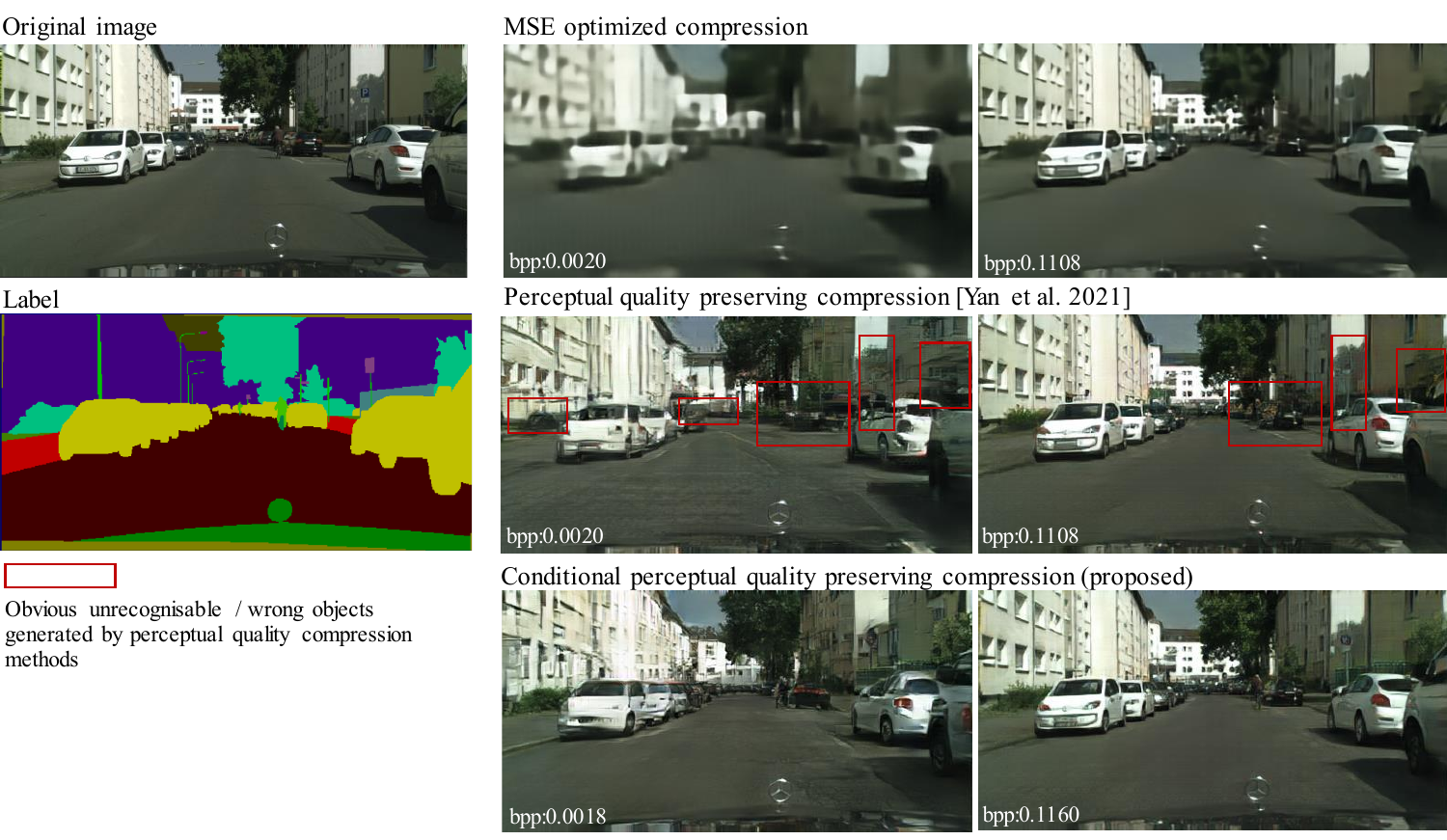}
 \caption{Qualitative comparison of different methods on Cityscape dataset. We mark obvious unrecognisable / wrong objects of two perceptual quality preserving methods in red. }
 \label{fig:cspeg}
\end{figure}

\textbf{Qualitative Results} We show the qualitative comparison of different methods in Fig.~\ref{fig:cspeg}. Obviously the MSE-optimized image is blurry and un-recognisable at low bpp. Though the perceptual codec \citep{Yan2021OnPL} is visually pleasing, it generates wrong objects. And our proposed approach leads to visually pleasing image with accurate semantic information at the same time.

\section{Related Work}
The perceptual quality by \citet{blau2018perception} and rate-distortion-perception trade-off \citep{,blau2019rethinking} have been well discussed in previous section. Another work related to us is the classification-distortion-perception trade-off \citep{liu2019classification}. It considers $Y$, the classification label as a separate term while we consider the posterior of image on $Y$. On the other hand, recent studies in information theory \citet{wagner2022rate,chen2022rate} conclude that no deterministic lossy codec exists for strong-sense perfect perceptual quality, which is different from our definition.

The pioneers of perceptual quality preserving image compression\citep{Rippel2017RealTimeAI,tschannen2018deep} achieve perfect perceptual quality using generative model before perception-distortion trade-off \citep{blau2018perception}. Later works \citep{Mentzer2020HighFidelityGI} improve the usability by incorporating extra distortion constrains. \citep{Zhang2021UniversalRR,Yan2021OnPL} show that we can achieve any perception-distortion trade-off without changing bitstream. Furthermore, \citet{Freirich2021ATO,Yan2022OptimallyCP} show that perception-distortion trade-off can be achieved by linear interpolation between images, without complicated decoder manipulation \citep{Iwai2020FidelityControllableEI,Agustsson2022MultiRealismIC}. There are other works utilize segmentation map for perceptual quality that resemble the experiment in Section 4.2 \citep{Agustsson2018ExtremeLI,agustsson2019generative,duan2022jpd}. However, they are limited to $Y$ as segmentation map and do not have theoretical optimality.


\section{Discussion \& Conclusion}

One limitation is that the current $Y$ chosen for experiment remains simple. It would be more interesting to see the results when $Y$ is image caption or description, combining the text-to-image models such as DALL-E \citep{pmlr-v139-ramesh21a}. Furthermore, the image resolution and model size used in this paper is relatively small. It would be more interesting to see the results on large size image and large model.

To conclude, we propose conditional perceptual quality, an extension of perceptual quality proposed by \citet{blau2018perception}. We show that it shares similar theoretical properties as rate-distortion-perception trade-off \citep{blau2019rethinking}. Based on those properties, we propose an optimal conditional perceptual quality preserving codec. And we empirically show that our codec successfully achieves high perceptual quality and semantic quality at the same time. Furthermore, we settle the argument about the randomness requirement to achieve perfect (conditional) perceptual quality by providing a lowerbound on randomness required.

\bibliography{ref_1}


\newpage

\appendix

\section{Proof of Main Results}
\label{app:proof}

In this section we give the the proof of main theoretical results.

\textbf{Theorem 2.1.}\textit{
(monotonicity and convexity) $R^I_C(D,P)$ is monotonously non-increasing in $D,P$. And when $d(.,.)$ is convex in its second argument, $R^I_C(D,P)$ is convex.
}
\begin{proof}
    The proof of monotonicity and convexity largely follows \citet{blau2019rethinking}. 
    
    The feasible region for $p(\hat{X}|X,Y)$ increases as $D,P$ increases. Therefore for $D_0\le D_1,P_0\le P_1$, any feasible $p(\hat{X}|X,Y)$ for $D_0,P_0$ is also feasible for $D_1,P_1$. And therefore, $R_C^I(D,P)$ is non-increasing in $D,P$.

    To show convexity of $R_C^I(D,P)$, we need to show for any convex combination we have:
    \begin{align}
        R^I_C(\lambda D_1+(1-\lambda)D_2,\lambda P_1+(1-\lambda)P_2) \le \lambda R^I_C(D_1,P_1)+(1-\lambda) R^I_C(D_2,P_2)
    \end{align}
    We prove this by constructing $p_{\hat{X}_{\lambda}|X,Y} = \lambda p_{\hat{X}_1|X,Y} + (1-\lambda) p_{\hat{X}_2|X,Y}$, where $p_{\hat{X}_1|X,Y},p_{\hat{X}_2|X,Y}$ are optimal solution to $R_C^I(D_1,P_1),R_C^I(D_2,P_2)$.

    Consider the convex combination of mutual information. As $I(X;\hat{X})$ is convex in $p_{\hat{X}|X}$, we have:
    \begin{align}
        I(X;\hat{X}_{\lambda}|Y) \le \lambda I(X;\hat{X}_1|Y) + (1-\lambda) I(X;\hat{X}_2|Y)
    \end{align}
    Consider the convex combination of $d(p_{X|Y},p_{\hat{X}_1|Y})\le P_1,d(p_{X|Y},p_{\hat{X}_2|Y})\le P_2$, as $d(.,.)$ is convex in second argument, we have:
    \begin{align}
        d(p_{X|Y},p_{\hat{X}_{\lambda}|Y}) &\le \lambda d(p_{X|Y},p_{\hat{X}_1|Y}) + (1-\lambda) d(p_{X|Y},p_{\hat{X}_2|Y}) \notag \\
        &\le \lambda P_1+(1-\lambda)P_2
    \end{align}
    Similarly, as the distortion is a linear function in distribution:
    \begin{align}
        \mathbb{E}[\Delta(X,\hat{X}_{\lambda})|Y] &= \lambda \mathbb{E}[\Delta(X,\hat{X}_1)|Y] + (1-\lambda) \mathbb{E}[\Delta(X,\hat{X}_2)|Y] \notag \\
        &\le \lambda D_1+(1-\lambda)D_2
    \end{align}
    Then by definition of $R_C^I(D,P)$, $p_{\hat{X}_{\lambda}|X,Y}$ is feasible to $R^I_C(\lambda D_1+(1-\lambda)D_2,\lambda P_1+(1-\lambda)P_2)$, and we have:
    \begin{align}
       R^I_C(\lambda D_1+(1-\lambda)D_2,\lambda P_1+(1-\lambda)P_2) &\le I(X;\hat{X}_{\lambda}|Y) \notag \\
       &\le \lambda I(X;\hat{X}_1|Y) + (1-\lambda) I(X;\hat{X}_2|Y) \notag \\
       &=\lambda R_C^I(D_1,P_ 1) + (1-\lambda) R_C^I(D_2,P_2)
    \end{align}
    And this completes the proof.
\end{proof}

\textbf{Theorem 2.2.}\textit{
(one-shot achievability) There exists a one-shot codec $f^0(.,Y),g^0(.,Y)$ that satisfies $\mathbb{E}[\Delta(X,\hat{X})|Y]\le D,d(p_{X|Y},p_{\hat{X}|Y})\le P$ if $R^{0}> R^I_C(D,P) + \log (R^I_C(D,P) + 1) + 5$;
}
\begin{proof}
    The proof of Theorem 2.2 and 2.3 is based on the conditional extension of strong functional representation lemma (SFRL) \citep{li2018strong}. The general framework follows the channel simulation proof by \citet{li2018strong} and achievability proof by \citet{theis2021coding}.  As the randomness is presented and we can not simply use joint asymptotic equipartition property \citep{cover1999elements}.

     By definition of $R^I_C(D,P)$, $\forall \epsilon>0$, there exist a $p(\hat{X}|X,Y)$ that satisfies the constrains $\mathbb{E}[\Delta(X,\hat{X})|Y] \le D,d(p_{X|Y},p_{\hat{X}|Y})\le P$, and:
    \begin{align}
        I(X;\hat{X}|Y) \le R^I_C(D,P) + \epsilon
    \end{align}
    By conditional extension of SFRL \citep{li2018strong}, there exist a $Z\perp (X,Y)$ and can represent $\hat{X}=q(X,Y,Z)$, where $q(.,.,.)$ is the quantization mapping, and 
    \begin{align}
        H(\hat{X}|Z,Y) \le I(X;\hat{X}|Y) + \log (I(X;\hat{X}|Y)+1) + 4
    \end{align}
    
    We assume the encoder and decoder share an unlimited randomness $W$. Let $Z=W$ be the shared randomness, upon observing $X$, the encoder side compute $\hat{X}=q(X,Y,Z)$ and encode $\hat{X}$ under the condition of $Y$ and $Z$. Then from Kraft inequality \citep{cover1999elements}, there exist a codec that can encode $\hat{X}$ into uniquely decodable bitstream $M$. Now the total expected code length:
    \begin{align}
        R^0 &= \mathbb{E}[\mathcal{L}(M)] \notag \\
        &\le H(\hat{X}|Y,Z) + 1 \notag \\
        & \le I(X;\hat{X}|Y) + \log (I(X;\hat{X}|Y) + 1) + 5 \notag \\
        & \le R^I_C(D,P) + \log(R^I_C(D,P)+1 + \epsilon) + 5 + \epsilon \notag \\
        & \le R^I_C(D,P) + \log(R^I_C(D,P)+1) + 5 + 2\epsilon
    \end{align}
    And this completes the proof.
\end{proof}
\textbf{Theorem 2.3.} \textit{
(achievability and converse) When $n\rightarrow\infty,$, there exists a codec $f^n(.,Y),g^n(.,Y)$ that satisfies $\mathbb{E}[\Delta(X^n,\hat{X}^n)|Y]\le D,d(p_{X|Y},p_{\hat{X}|Y})\le P$ if $R^{n}<R^{I}_C(D,P)$ and no codec with $R^{n}<R^{I}_C(D,P)$;
}
\begin{proof}
    We first proof the achievability.

    By definition of $R^I_C(D,P)$, $\forall \epsilon>0$, there exist a $p(\hat{X}|X,Y)$ that satisfies the constrains $\mathbb{E}[\Delta(X,\hat{X})|Y] \le D,d(p_{X|Y},p_{\hat{X}|Y})\le P$, and:
    \begin{align}
        I(X;\hat{X}|Y) \le R^I_C(D,P) + \epsilon_1
    \end{align}
    As the proof of Theorem 2.2, we have a shared randomness $Z$ that satisfies:
    \begin{align}
        \frac{1}{n}H(\hat{X}^n|Y^n,Z) &\le \frac{1}{n}(I(X^n;\hat{X}^n|Y^n) + \log (I(X^n;\hat{X}^n|Y^n)+1) + 4) \notag \\
        & =  I(X;\hat{X}|Y) + \frac{1}{n}(\log (nI(X;\hat{X}|Y)+1) + 4) \notag \\
        & \le R_C^I(D,P) + \frac{1}{n}(\log (nR_C^I(D,P)+2) + 5)
    \end{align}
    Then from the theory of typical set \citep{cover1999elements}, we can construct a codebook with size $2^{H(\hat{X}^n|Y^n)+\epsilon_2}$, which means that we can achieve a rate of $R_C^I(D,P) + \epsilon_3$. And this completes the proof of achievability.

    The proof of converse is relatively simple. It closely follows the proof of converse of rate distortion by \citet{cover1999elements}:
    \begin{align}
        nR^n &\ge H(f^n(X^n)|Y) \notag \\
        & \ge I(X^n;f^n(X^n)|Y) \notag \\
        & \overset{(a)}{\ge} I(X^n;\hat{X}^n|Y) \notag\\
        & = H(X^n|Y) - H(X^n|\hat{X}^n,Y) \notag\\
        & \ge \sum_{i=1}^n (H(X_i|Y) - H(X_i|\hat{X}_i,Y)) \notag\\
        & = \sum_{i=1}^n I(X_i;\hat{X}_i|Y) \notag\\
        & \overset{(b)}{\ge} \sum_{i=1}^n R^I_C(\mathbb{E}[\Delta(X_i,\hat{X}_i)|Y],d(p_{X_i|Y},p_{\hat{X}_i|Y})) \notag\\
        & \overset{(c)}{\ge} n R^I_C(\mathbb{E}[\sum_{i=1}^n\Delta(X_i,\hat{X}_i)|Y],\sum_{i=1}^n d(p_{X_i|Y},p_{\hat{X}_i|Y})) \notag\\
        & = n R^I_C(\mathbb{E}[\Delta(X^n,\hat{X}^n)|Y], d(p_{X|Y},p_{\hat{X}|Y})) \notag\\
        & \overset{(d)}{\ge} nR^I_C(D,P),
    \end{align}
    where (a) is due to data pre-processing inequality \citep{cover1999elements}, (b) is the definition of $R^I_C(D,P)$, (c) is due to the convexity of $R_C^I(D,P)$ in Theorem 2.1, (d) is due to the monotoncity of $R_C^I(D,P)$ in Theorem 2.1.
\end{proof}

\textbf{Theorem 2.4.}\textit{
Perfect conditional perceptual quality leads to perfect perceptual quality.
}
\begin{proof}
By definition of divergence, $d(p,q)=0$ if and only if $p=q$. Therefore, $d(p_{X|Y},p_{\hat{X}|Y})=0$ leads to $p_{X|Y}=p_{\hat{X}|Y}$, which further leads to $\mathbb{E}_{p_Y}[p_{X|Y}]=\mathbb{E}_{p_Y}[p_{\hat{X}|Y}]=p_X=p_{\hat{X}}$. And finally, we have $p_X=p_{\hat{X}}$, which implies perfect perceptual quality.
\end{proof}
\textbf{Theorem 2.5.}\textit{
$R^{I}_C(D,0) \le R^{I}_C(\frac{1}{2}D,\infty)$;
}
\begin{proof}
    This proof closely follows the proof of Theorem. 2 by \citet{blau2019rethinking}. Specifically, we first compute the posterior of $X$ given reconstruction $\hat{X}$:
    \begin{align}
        p_{X|\hat{X},Y}=\frac{p_{\hat{X}|X,Y}p_{X|Y}}{p_{\hat{X}|Y}}
    \end{align}
    Then, we directly add a posterior mapping from $\hat{X}$ to $\tilde{X}$ using the true posterior:
    \begin{align}
        p_{\tilde{X}|\hat{X},Y} = p_{X|\hat{X},Y}
    \end{align}
    Then it is obvious that the joint distribution is also the same: $p_{\tilde{X},\hat{X}|Y} = p_{X,\hat{X}|Y}$. And therefore, the marginal distribution is also the same: $p_{\tilde{X}|Y}=p_{X|Y}$. And therefore, the post-processing mapping $p_{\tilde{X}|\hat{X},Y}$ achieves perfect perceptual quality. 
    Next, we show the MSE distortion of $\tilde{X}$ is bounded:
    \begin{align}
        \mathbb{E}[||X-\tilde{X}||^2|Y] &= \mathbb{E}[||X||^2|Y] - 2\mathbb{E}[X^T\tilde{X}|Y] + \mathbb{E}[||\tilde{X}^2|||Y] \notag \\
        &\overset{(a)}{=} 2\mathbb{E}[||X||^2|Y] - 2\mathbb{E}[X^T\tilde{X}|Y] \notag  \\
        &= 2\mathbb{E}[||X||^2|Y] - 2\mathbb{E}_{p_{\hat{X}|Y}}[\mathbb{E}_{p_{X,\tilde{X}|\hat{X},Y}}[X^T\tilde{X}|\hat{X},Y]|Y] \notag \\
        &\overset{(b)}{=} 2\mathbb{E}[||X||^2|Y] - 2\mathbb{E}_{p_{\hat{X}|Y}}[||\mathbb{E}_{p_{X|\hat{X},Y}}[X|\hat{X},Y]||^2|Y] \notag \\
        &\overset{(c)}{=} 2\mathbb{E}[||X||^2|Y] - 2\mathbb{E}_{p_{\hat{X}|Y}}[||\hat{X}||^2|Y] \notag \\
        &\overset{(d)}{=} 2\mathbb{E}[||X-\hat{X}||^2|Y] \notag \\ 
    \end{align}
    where in (a) we use $p_{X}=p_{\tilde{X}}$, in (b) we use the fact that $X,\tilde{X}$ is conditional independent given $\hat{X},Y$, in (c) we use $\hat{X}=\mathbb{E}[X|\hat{X},Y]$, as $\mathbb{E}[X|\hat{X},Y]$ is the optimal solution to minimize MSE. And (d) is equivalent to $\mathbb{E}[X^T\hat{X}|Y] = \mathbb{E}[||\hat{X}||^2|Y]$. And this can be shown by
    \begin{align}
    \mathbb{E}[X^T\hat{X}|Y] &= \mathbb{E}[\mathbb{E}[X^T\hat{X}|\hat{X},Y]|Y] \notag \\
    &\overset{(e)}{=} \mathbb{E}[\mathbb{E}[\mathbb{E}[X|\hat{X},Y]^T\mathbb{E}[\hat{X}|\hat{X},Y]|\hat{X},Y]|Y] \notag \\
    & \overset{(f)}{=} \mathbb{E}[\mathbb{E}[\hat{X}^T\hat{X}]|Y] \notag \\
    & = \mathbb{E}[||\hat{X}||^2|Y],
    \end{align}
    where (e) is due to the fact that $X,\hat{X}$ are conditional independent given $\hat{X}$ and (f) uses $\hat{X}=\mathbb{E}[X|\hat{X},Y]$, as $\mathbb{E}[X|\hat{X},Y]$ is the optimal solution to minimize MSE. And this completes the proof.
\end{proof}

To proof Theorem 2.6, we need the following lemma:

\textbf{Lemma 2.6.1.}\textit{
For optimal one-shot codec, we have deterministic decoder $\hat{X}=\mathbb{E}[X|M,Y]$. On the other hand, for any $\hat{X}$, there exists only one $M$ that decodes into $\hat{X}$. To conclude, the optimal one-shot decoder is a deterministic invertible mapping from $M$ to $\hat{X}$.}
\begin{proof}
    As $M^0$ is given, the rate is fixed. For any $\hat{X}\neq\mathbb{E}[X|M^0,Y]$, we just replace it by $\mathbb{E}[X|M^0,Y]$. Then obviously, the distortion is lower, and that is paradox to the optimality of codec. Therefore, we must have $\hat{X}=\mathbb{E}[X|M^0,Y]$.

    Similarly, assume both $M^0_1$ and $M^0_2$ decode into the same $\hat{X}$. Then obviously the bitrate to distinguish between $M^0_1$ and $M^0_2$ is redundant. More specifically, given $\hat{X}$, the bitrate has $-\log p(M_1^0) - \log p(M_2^0) + \log (p(M_1^0)+p(M_2^0))$ to improve. And this is paradox to the optimality of codec. Therefore, one $\hat{X}$ can only have one corresponding $M$.
\end{proof}
\textbf{Theorem 2.6.}\textit{
Given an optimal one-shot codec $f^0(.,Y),g^0_1(.,Y)$ with code $M^0$, reconstruction $\hat{X}$ and distortion $\mathbb{E}[\Delta(X,\hat{X})|Y]\le D/2$. there exists an optimal perceptual decoder $g_2(.,Y)$ with $p_{\tilde{X}|M^0,Y}= p_{X|M^0,Y}, d(p_{X|Y},p_{\tilde{X}|Y})=0,\mathbb{E}[\Delta(X,\tilde{X})|Y]\le D$;}
\begin{proof}
    To proof Theorem 2.6, we first prove that their exist a decoder with decoding function $p_{\tilde{X}|M^0,Y}= p_{X|M^0,Y}$, perfect perceptual quality $d(p_{X|Y},p_{\tilde{X}|Y})=0$, and distortion $\mathbb{E}[\Delta(X,\tilde{X})|Y]\le D$.
    
    As Lemma 2.6.1 tells us that $\hat{X}$ and $M$ are deterministic invertible mapping, then we have $p_{X|M,Y}=p_{X|\hat{X},Y}$ and $p_{\tilde{X}|M,Y}=p_{\tilde{X}|\hat{X},Y}$. And the rest of proof follows the proof of Theorem 2.5. Similar trick can also provide an alternative proof of Theorem. 2 in \citet{Yan2021OnPL}, which is the non-conditional version of Theorem 2.6.

    Next, we prove the optimality of such decoder. Consider for each $y\in \mathcal{Y}$, we iteratively apply Theorem. 1 of \citet{Yan2021OnPL} and construct $|\mathcal{Y}|$ decoders. According to \citet{Yan2021OnPL}, those decoders are optimal for each $y$ and satisfies $p_{X|M,Y=y}=p_{X|\hat{X},Y=y}$. Next, let's construct a meta-decoder by combining those optimal decoders, and we notice that the distortion and perception are satisfied. Furthermore, we have $p_{X|M,Y}=p_{X|\hat{X},Y}$. And such meta-decoder is a valid solution to the constrains. We notice that the rate of this decoder is a linear combination of rate for each $y$. Let's assume another codec with rate less than this rate exists, then for at least one of $y\in\mathcal{Y}$, the rate of this codec is less than the optimal rate by Theorem. 1 of \citet{Yan2021OnPL}. And this brings paradox. Thus, no other codec satisfying the distortion perception constrains has strictly less rate than the meta-decoder. And we can say the meta-decoder with $p_{X|M,Y}=p_{X|\hat{X},Y}$ is optimal.
\end{proof}

\textbf{Theorem 3.}\textit{
Assume $d(p_{X|Y},p_{\hat{X}|Y})=0$ and $Y$ is deterministic of $X$, for any codec with total expected code length $R$ and code $M$, there exists another codec that has the same code $M$ following the above framework with $Y$ encoded losslessly, whose total expected code length $R_M+R_Y \le R+2$.
}

\begin{proof}
Denote the encoder of first codec as $f(.,Y)$, and the code as $M=f(X,Y)$. For the second codec, we do not change encoder or decoder. Instead, as both the encoder and decoder has $Y$ before we encode and decode $M$, we encode $M$ with conditional entropy model with entropy $H(M|Y)$. And the total entropy for the second codec is:
\begin{align}
    H(M,Y) & = H(M|Y) + H(Y) \notag \\
    & = H(M) + H(Y|M) \label{eq:mjy}
\end{align}
We note that the first line of Eq.~\ref{eq:mjy} is the entropy of the second codec, and the second line is the entropy of first codec plus an overhead $H(Y|M)$. Next, we examine this overhead $H(Y|M)$:
\begin{align}
    H(Y|M) & \overset{(a)}{\le} H(Y|\hat{X}) \notag \\
    & \overset{(b)}{=} H(Y|X) \notag \\
    & \overset{(c)}{=} 0
\end{align}
where in (a) we use data processing inequality \citep{cover1999elements}, in (b) we use the assumption that $d(p_{X|Y},p_{X|\hat{Y}})=0$ and in (c) we use the assumption that $Y$ is deterministic of $X$. Therefore, we have $H(M|Y)+H(Y) = H(M)$.

From Kraft's inequality \citep{cover1999elements}. 
, we have:
\begin{align}
    H(Y) &\le R_Y\le H(Y)+1 \notag \\
    H(M|Y) &\le R_M\le H(M|Y) + 1 \notag \\
    H(M) &\le R \le H(M)+1
\end{align}
Therefore, we have:
\begin{align}
    R_Y+R_M&\le H(Y)+H(M|Y)+2 \notag \\
           & \overset{(d)}{=} H(M) + 2 \notag \\
           & \le R + 2
\end{align}
, where (d) use previous result that $H(Y|M) = 0$. And this completes the proof.
\end{proof}
\textbf{Theorem 4.}\textit{
For one-shot codec, to achieve perfect perceptual quality at rate $R^0$, we require a common randomness $W \in \mathcal{W}$ with $\log |\mathcal{W}| \ge H(W) \ge H(X) - (R^0+1)$. To achieve perfect conditional perceptual quality, we require $\log |\mathcal{W}| \ge H(W) \ge H(X|Y) - (R^0+1)$.
}
\begin{proof}
    Let's first consider the perfect perceptual quality case. Consider a one-shot codec encoding image $X$ into bitstream $M$, and reconstructs $\hat{X}$, we have:
    \begin{align}
        H(\hat{X},M) &= H(\hat{X}|M) + H(M) \notag \\
        &= H(\hat{X})+H(M|\hat{X}) \notag \\
        &\ge H(\hat{X}) \notag \\
        &\overset{(a)}{=} H(X),
    \end{align}
    where (a) is due to $d(p_X,p_{\hat{X}})=0$. And this leads to:
    \begin{align}
        H(\hat{X}|M) &\ge H(X) -H(M) \notag \\
        &\overset{(b)}{>} H(X) - (R^0 + 1),
    \end{align}
    where (b) is due to Kraft inequality \citep{cover1999elements}. For lossy compression, as $H(\hat{X}|M)>0$, we need a stochastic decoder. Specifically, given code $M$, the decoder has a reconstruction $\hat{X}$ whose conditional entropy is at least $H(\hat{X}|M)$. When decoder is a conditional GAN, this means that the noise $W$'s entropy is at least $H(\hat{X}|M)$. To achieve this, we consider a randomness $W \in \mathcal{W}$ shared by encoder and decoder, which is independent of $M$. With $W$, we require $\hat{X}$ to be deterministic, otherwise the encoder side has no idea of what the decoder side reconstruction $\hat{X}$ looks like. The coding procedure resembles the bits-back coding \citep{townsend2018practical}. When encoding, we first sample $\hat{X}$ from $p(\hat{X}|M)$ with an expected bitrate $H(\hat{X}|M)$. And this requires $W$ has a entropy of $H(\hat{X}|M)$. As the alphabet size of $W$, $|\mathcal{W}|$ is at least $2^{H(W)}$, we need an alphabet size $|\mathcal{W}|$ of at least $2^{H(\hat{X}|M)}$. During decoding, we also sample $\hat{X}$ from $p(\hat{X}|M)$ from the very same $W$. More specifically: 
    \begin{align}
        H(\hat{X}|W,M) &= 0 \notag \\ 
        & = H(\hat{X}|M) - H(W|M) \notag \\
        &\overset{(c)}{=} H(\hat{X}|M) - H(W),
    \end{align}
    where (c) is due to the fact that $W\perp M$. And from Theorem 2.6.4 of \citet{cover1999elements}, we have
    \begin{align}
        \log |\mathcal{W}| \ge H(W) \ge H(X) - (R^0+1).
    \end{align}
    Similar proof applies to the conditional perceptual quality case. We only add additional limitation that $W$ is also independent of $Y$ and condition $M,\hat{X}$ on $Y$, and everything else in the proof remains the same.
\end{proof}

\section{More Experimental Results}
\subsection{More Experiment Setup}
\label{app:eset}
\begin{figure}[htb]
\centering
 \includegraphics[width=\linewidth]{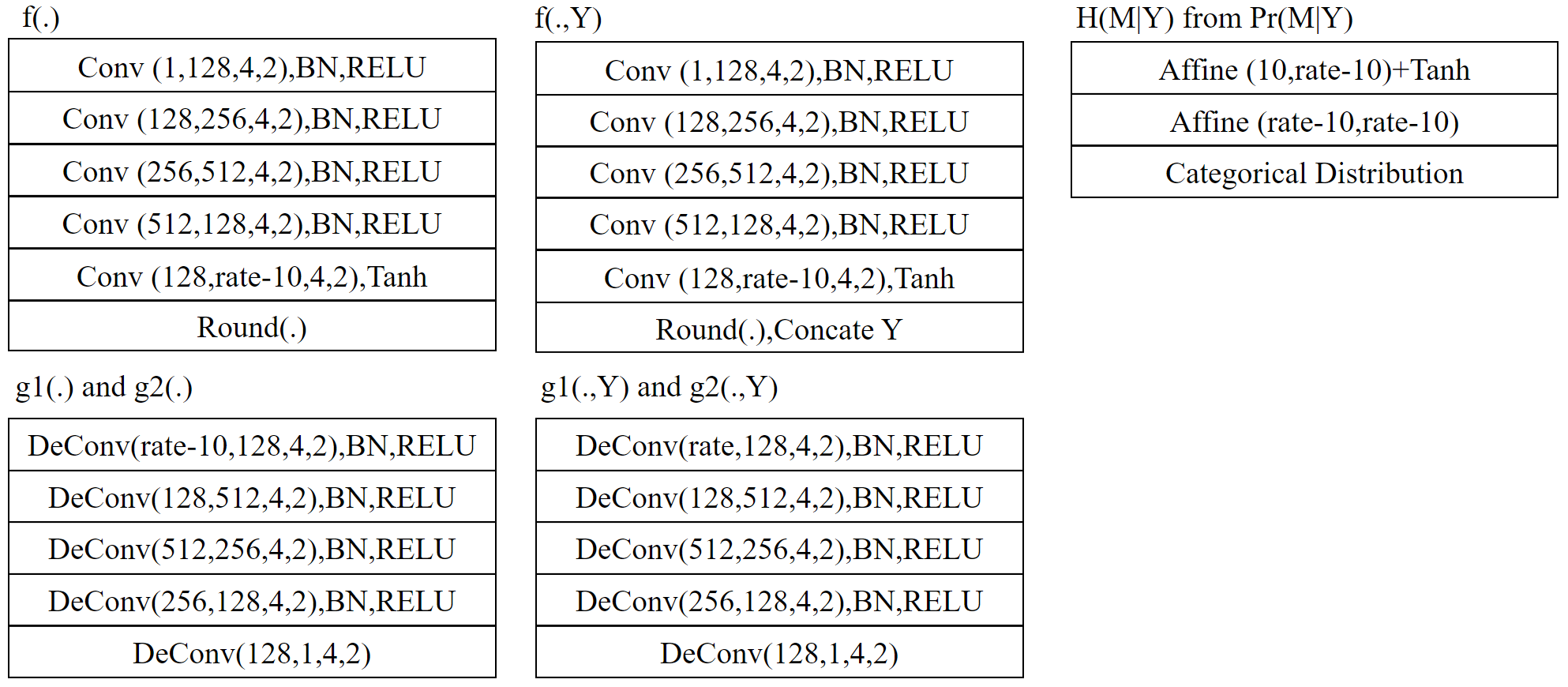}
 \caption{Detailed neural network architecture used in MNIST dataset. The naming of each components follows Fig.~\ref{fig:fw}. For Conv and Deconv layers, the parameters are input channels, output channels, kernel size and stride. For Affine layers, the parameters are input dimension and output dimensions. The rate parameter differs per bitrate settings. BN is a short for BatchNorm.}
 \label{fig:mnistarch}
\end{figure}

For experiment using MNIST dataset \citep{LeCun2005TheMD}, we use the default dataset split with 60000 training images and 10000 testing images. All the images are re-scaled into $32\times 32$ and no data augmentation is used. The encoder $f$, mse decoder $g_1$, perceptual decoder $g_2$ and discriminator $h$'s architecture follows \citep{Yan2021OnPL}. The optimizer is RMSprop \citep{hinton2012neural}. The learning rate is $10^{-3}$ for encoder, decoder $g_1$, decoder $g_2$ and $10^{-2}$ for discriminator $h$. The model is optimized for $100$ epochs with batch-size $128$. For the pre-trained classifier, we train a ResNet-18 \citep{He2015DeepRL} from scratch using the training dataset. The optimizer we adopt is stochastic gradient descent with momentum \citep{Qian1999OnTM}. The learning rate is $10^{-2}$ and the momentum is $0.9$. We optimize the model for $10$ epochs with batchsize $128$. The rate is directly set to $\{4,8,12,16,20\}$ by forcing the dimension of binary latent code as \citet{Yan2021OnPL}. In addition, we also present the detailed neural network architecture in Fig.~\ref{fig:mnistarch}.

\begin{figure}[htb]
\centering
 \includegraphics[width=\linewidth]{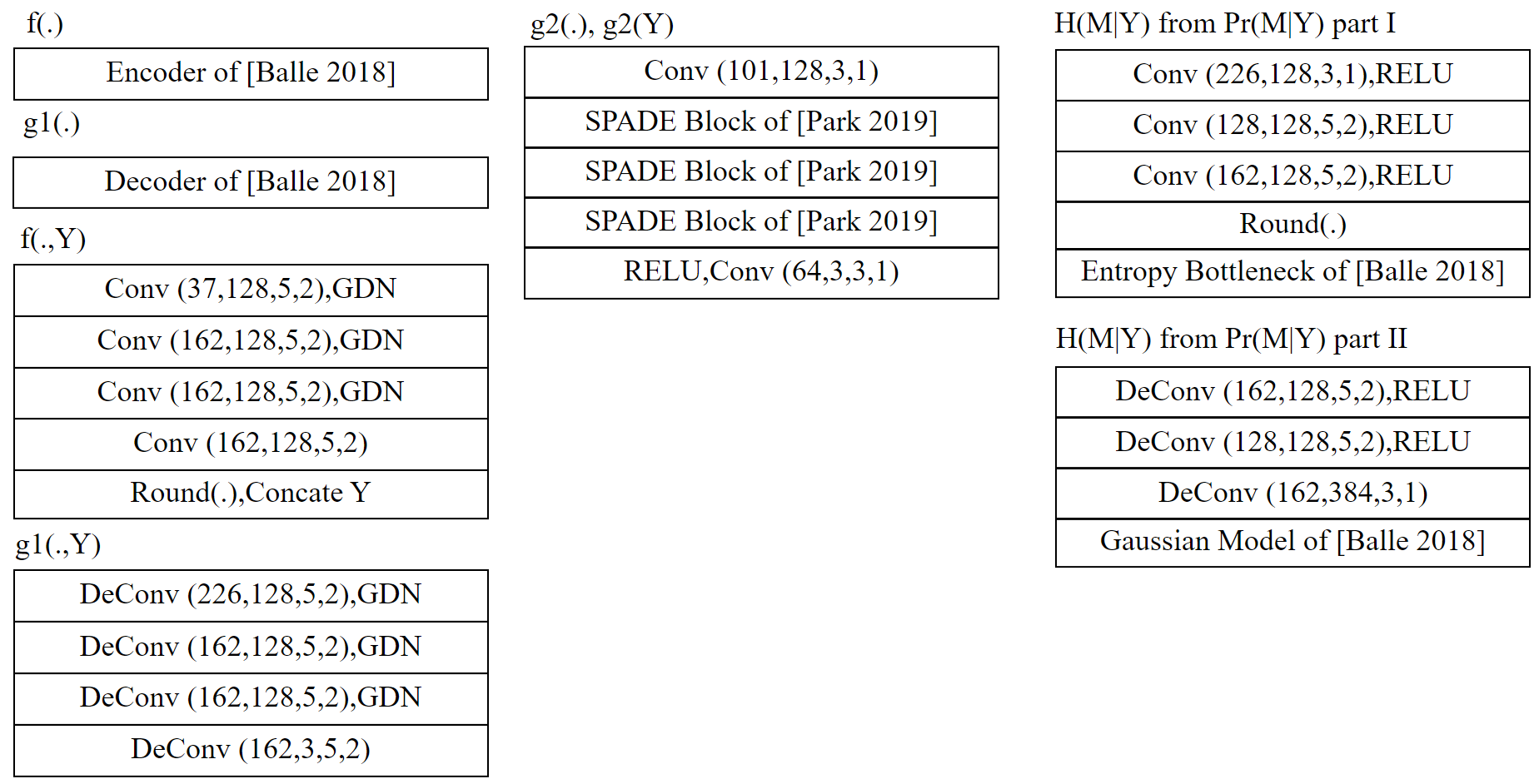}
 \caption{Detailed neural network architecture used in Cityscape dataset. We do not employ different architecture for $g_2(.)$ and $g_2(.,Y)$. Instead, we just set $Y$ to $0$ when we want $g_2(.)$.}
 \label{fig:csarch}
\end{figure}

For experiment using Cityscape dataset \citep{Cordts2016TheCD}, we use the default dataset split with 2975 training images and 500 validation images for test. All the images are re-scaled into $512\times 256$, and the random horizontal flip is adopted for data augmentation.  All the segmentation map is further re-scaled into $64\times 32$. The encoder $f$ and mse decoder $g_1$ follow the architecture of \citet{balle2018variational}. The perceptual generator $g_2$, discriminator $h$ follow the architecture of \citet{Sushko2020YouON}. We adopt Adam optimizer \citep{Kingma2014AdamAM}. The learning rate is $4\times10^{-3}$ for discriminator and $10^{-3}$ for other part of model. We optimize the model for $200$ epochs with batchsize $20$. The $\lambda$ trading off rate and distortion is $\{0.00015,0.0003,0.0006,0.00125,0.0025\}$. The neural network architecture used in this experiment is too complicated to be expanded exactly. Thus, we depicted the architecture in Fig.~\ref{fig:csarch} with modules copied from \citet{balle2018variational,park2019semantic}. We refer interested readers those papers or the source code for exact details.

\subsection{More Quantitative Results}
\label{app:quant}
In Section 4.2 we do not have room to list the bitrate and other metrics for distortion-conditional perception trade-off. In Tab.~\ref{tab:rdcpfull}, we provide all the metrics comparing linear interpolation of images and manual adjusting $\beta$.
\begin{table}[htb]
\centering
\caption{The rate, MSE, conditional FID, FID and accuracy of linear interpolation of images and manual adjusting $\beta$.}
 \label{tab:rdcpfull}
\begin{tabular}{@{}llllll@{}} \toprule
                                     & Rate (bits) $\downarrow$ & MSE $\downarrow$ & ConFID $\downarrow$ & FID $\downarrow$ & acc  $\uparrow$  \\ \midrule
\multirow{7}{*}{Image Interpolation} & 4.29        & 0.0821 & 25.55  & 14.33  & 0.9754 \\
                                     & 4.29        & 0.0741 & 27.22  & 15.96  & 0.9778 \\
                                     & 4.29        & 0.0668 & 33.78  & 20.72  & 0.9803 \\
                                     & 4.29        & 0.0553 & 55.07  & 38.15  & 0.9729 \\
                                     & 4.29        & 0.0469 & 89.15  & 65.69  & 0.9518 \\
                                     & 4.29        & 0.0422 & 136.14 & 104.01 & 0.9059 \\
                                     & 4.29        & 0.0405 & 225.54 & 178.80 & 0.8363 \\ \midrule
\multirow{5}{*}{$\beta$ Adjustment}  & 4.35        & 0.0783 & 39.29  & 25.32  & 0.9758 \\
                                     & 4.30        & 0.0760 & 46.69  & 31.61  & 0.9704 \\
                                     & 4.32        & 0.0598 & 62.66  & 43.68  & 0.9368 \\
                                     & 4.33        & 0.0507 & 82.14  & 57.95  & 0.9569 \\
                                     & 4.26        & 0.0431 & 112.51 & 83.10  & 0.9625 \\ 
                                     & 4.29        & 0.0405 & 225.54 & 178.80 & 0.8363 \\ \bottomrule
\end{tabular}
\end{table}

\subsection{More Qualitative Results}
\label{app:qualt}

\begin{figure}[htb]
\centering
 \includegraphics[width=1.0\linewidth]{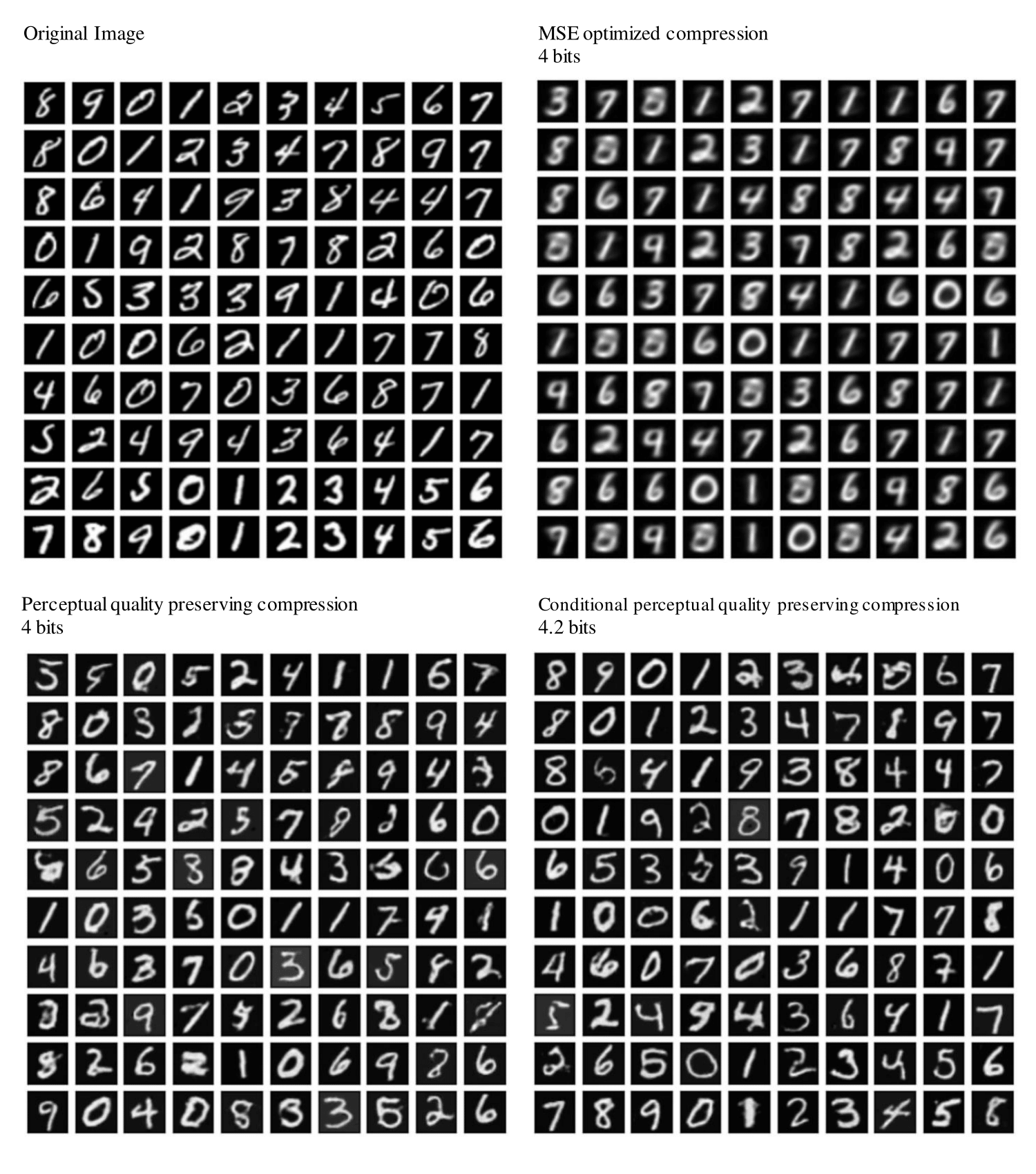}
 \caption{Qualitative results comparing original image, baseline MSE codec, perceptual quality preserving codec \citep{Yan2021OnPL} and our proposed conditional perceptual preserving codec for MNIST dataset with rate $\approx 4$ bits.}
 \label{fig:mnist1}
\end{figure}

\begin{figure}[htb]
\centering
 \includegraphics[width=1.0\linewidth]{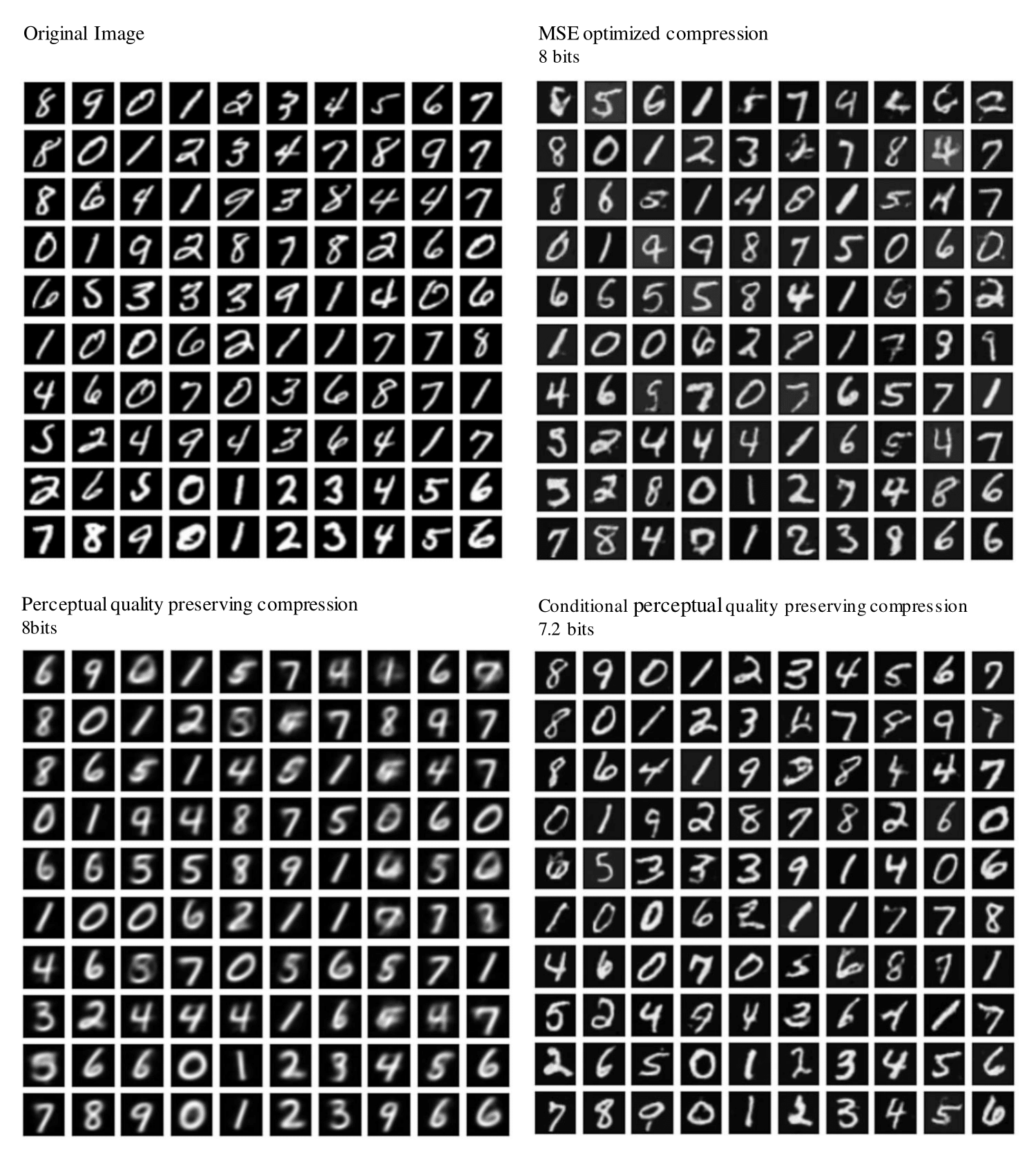}
 \caption{Qualitative results comparing original image, baseline MSE codec, perceptual quality preserving codec \citep{Yan2021OnPL} and our proposed conditional perceptual preserving codec for MNIST dataset with rate $\approx 8$ bits.}
 \label{fig:mnist2}
\end{figure}

\begin{figure}[htb]
\centering
 \includegraphics[width=1.0\linewidth]{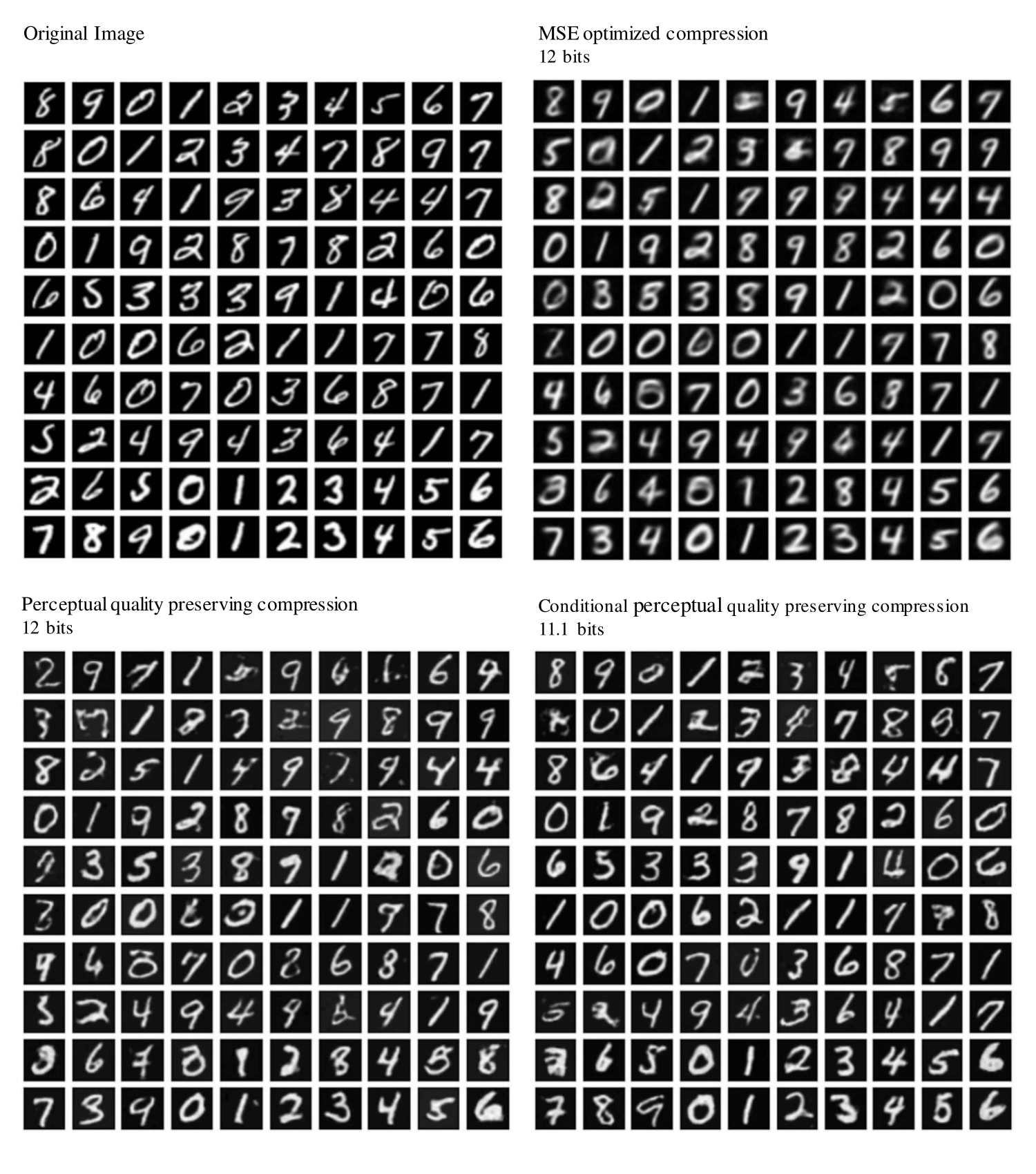}
 \caption{Qualitative results comparing original image, baseline MSE codec, perceptual quality preserving codec \citep{Yan2021OnPL} and our proposed conditional perceptual preserving codec for MNIST dataset with rate $\approx 12$ bits.}
 \label{fig:mnist3}
\end{figure}

\begin{figure}[htb]
\centering
 \includegraphics[width=1.0\linewidth]{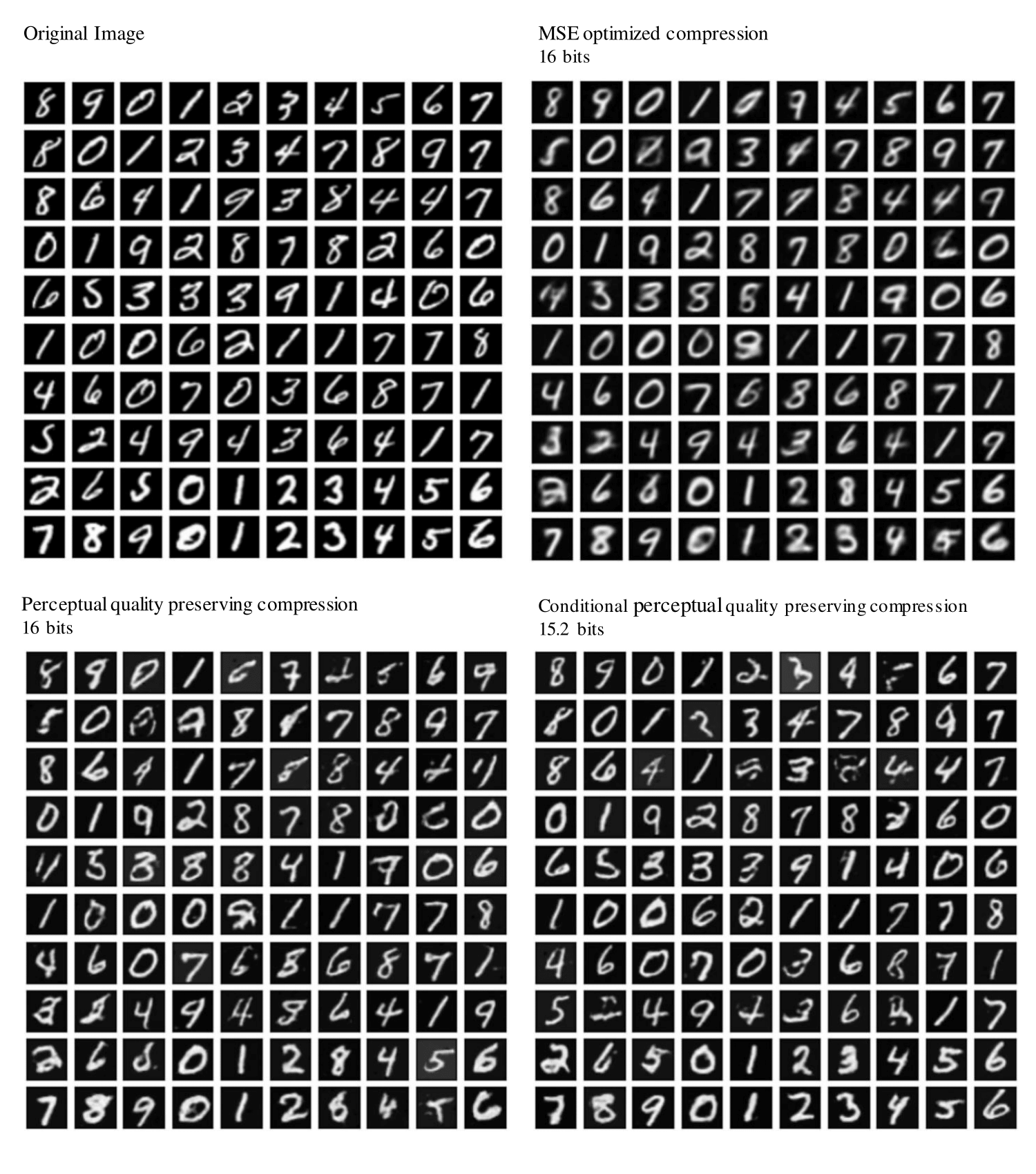}
 \caption{Qualitative results comparing original image, baseline MSE codec, perceptual quality preserving codec \citep{Yan2021OnPL} and our proposed conditional perceptual preserving codec for MNIST dataset with rate $\approx 16$ bits.}
 \label{fig:mnist4}
\end{figure}

\begin{figure}[htb]
\centering
 \includegraphics[width=1.0\linewidth]{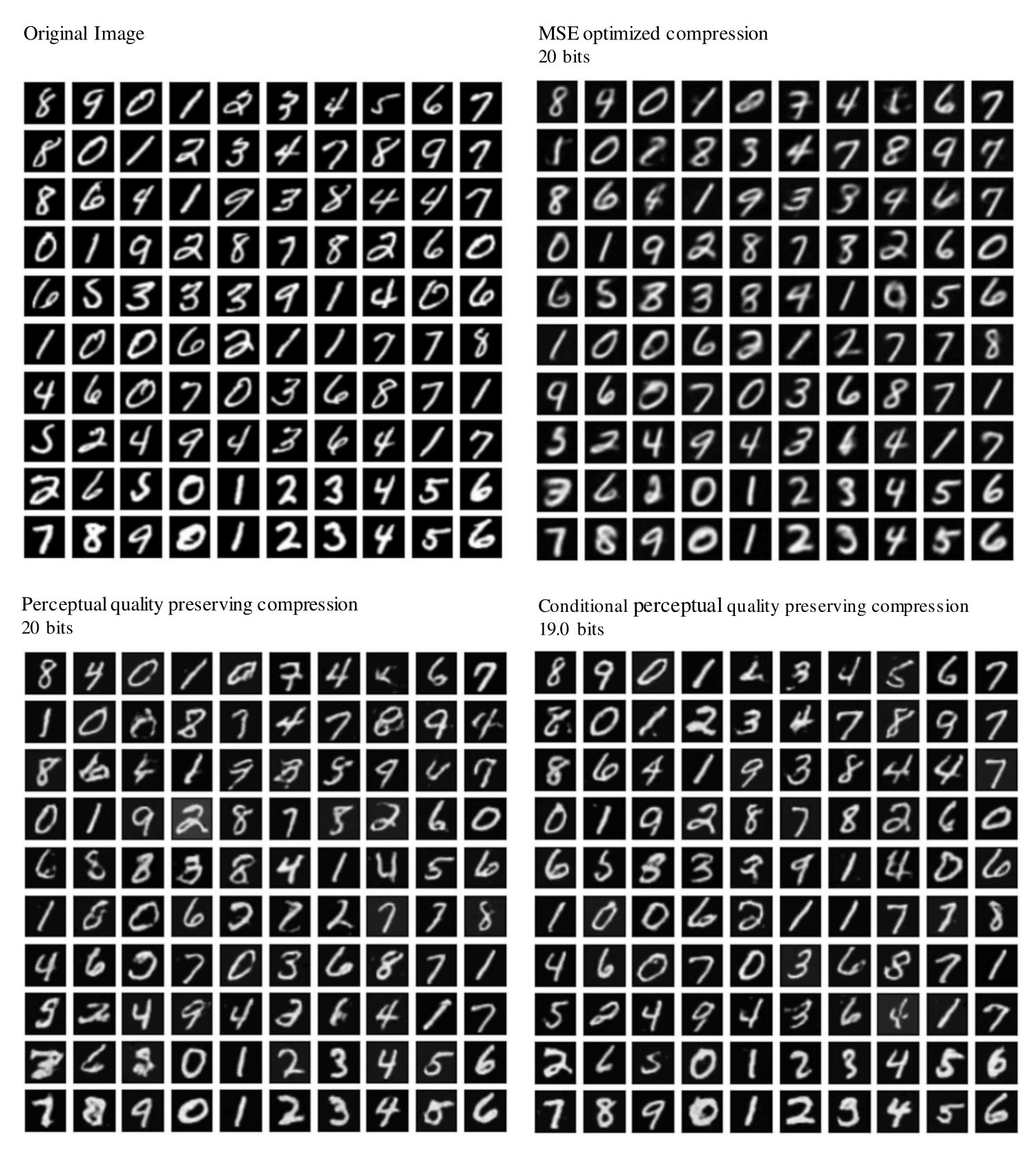}
 \caption{Qualitative results comparing original image, baseline MSE codec, perceptual quality preserving codec \citep{Yan2021OnPL} and our proposed conditional perceptual preserving codec for MNIST dataset with rate $\approx 20$ bits.}
 \label{fig:mnist5}
\end{figure}

\begin{figure}[htb]
\centering
 \includegraphics[width=1.0\linewidth]{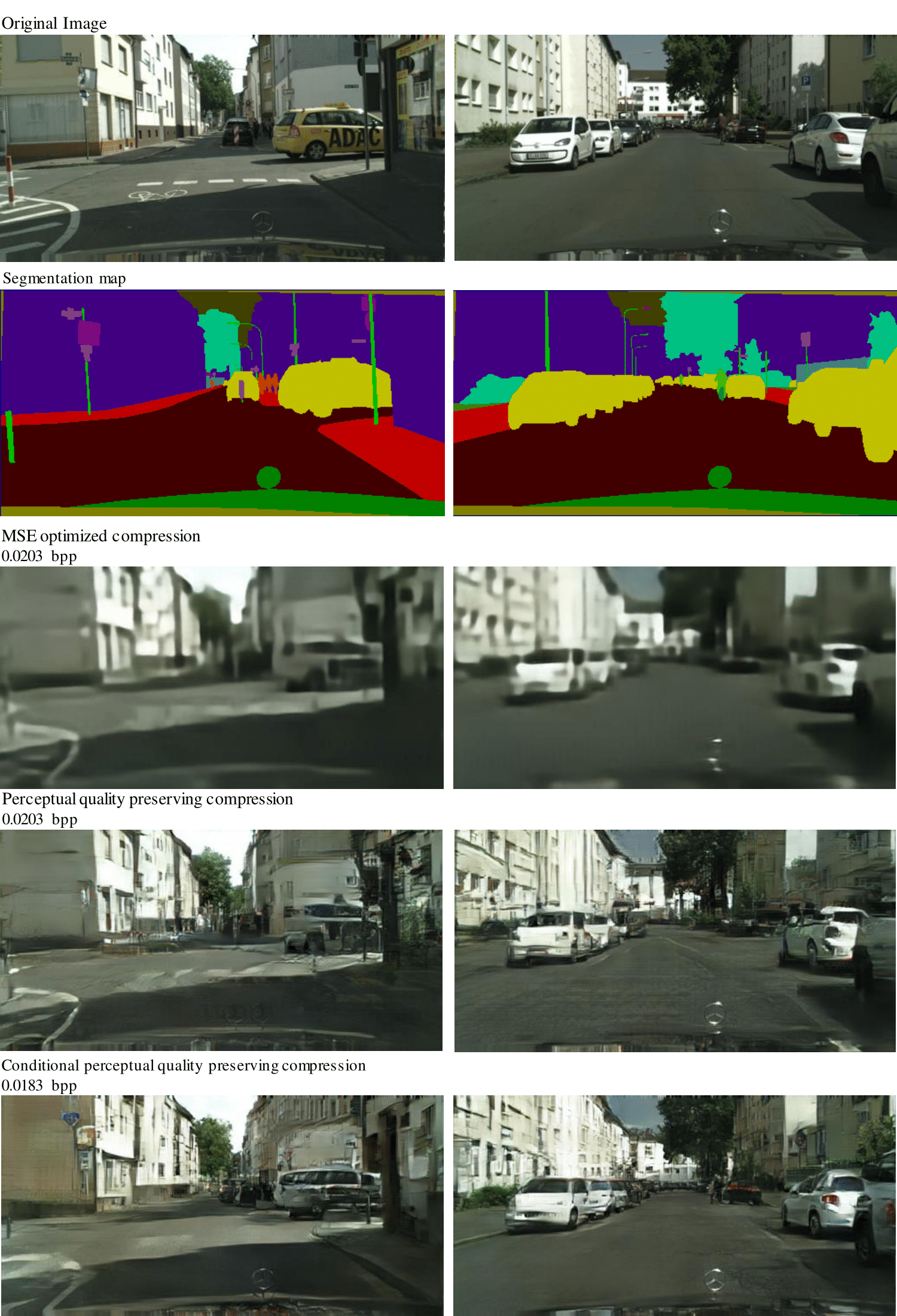}
 \caption{Qualitative results comparing original image, baseline MSE codec, perceptual quality preserving codec \citep{Yan2021OnPL} and our proposed conditional perceptual preserving codec for Cityscape dataset with rate $\approx 0.02$ bpp.}
 \label{fig:ctsp1}
\end{figure}

\begin{figure}[htb]
\centering
 \includegraphics[width=1.0\linewidth]{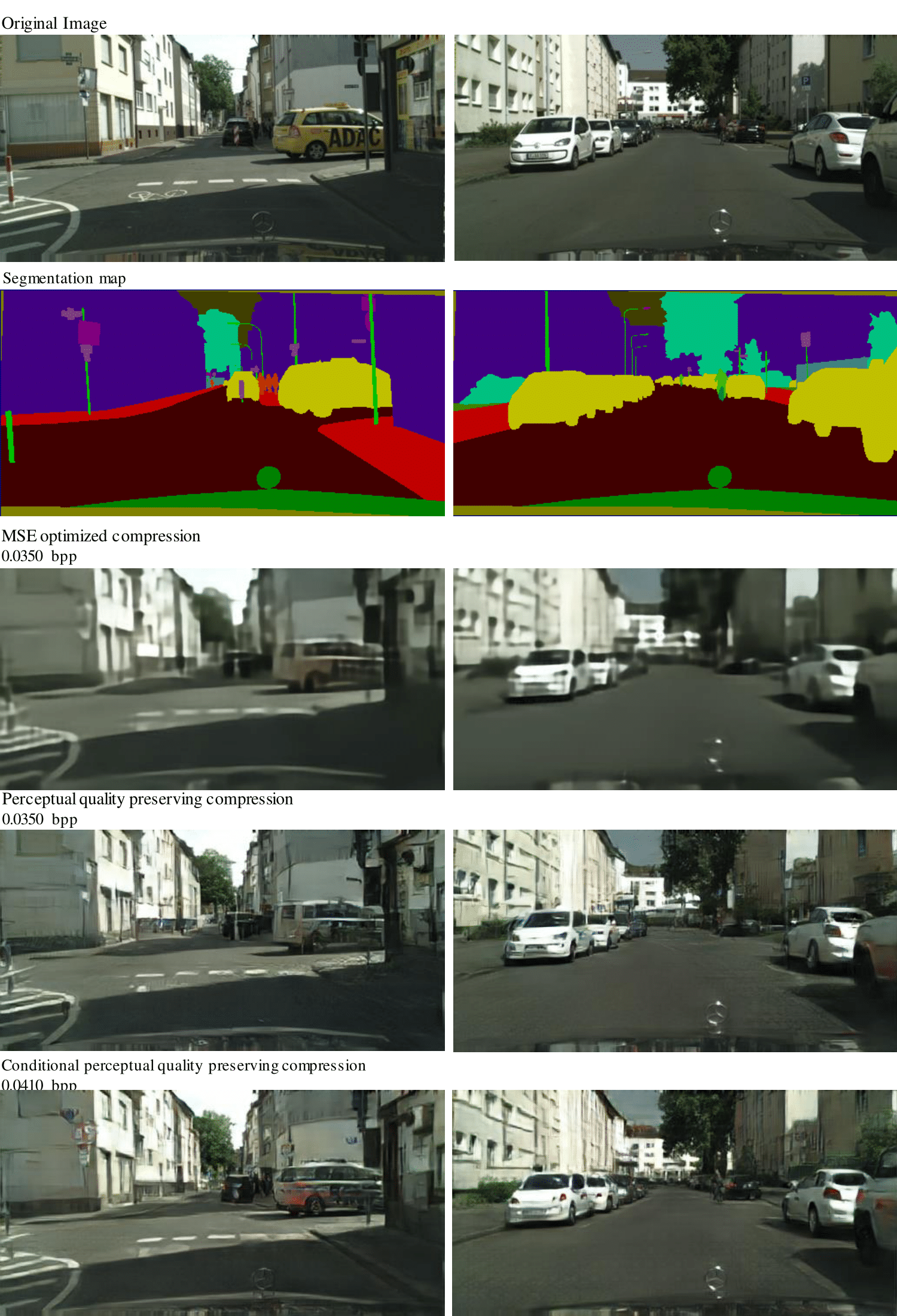}
 \caption{Qualitative results comparing original image, baseline MSE codec, perceptual quality preserving codec \citep{Yan2021OnPL} and our proposed conditional perceptual preserving codec for Cityscape dataset with rate $\approx 0.04$ bpp.}
 \label{fig:ctsp2}
\end{figure}

\begin{figure}[htb]
\centering
 \includegraphics[width=1.0\linewidth]{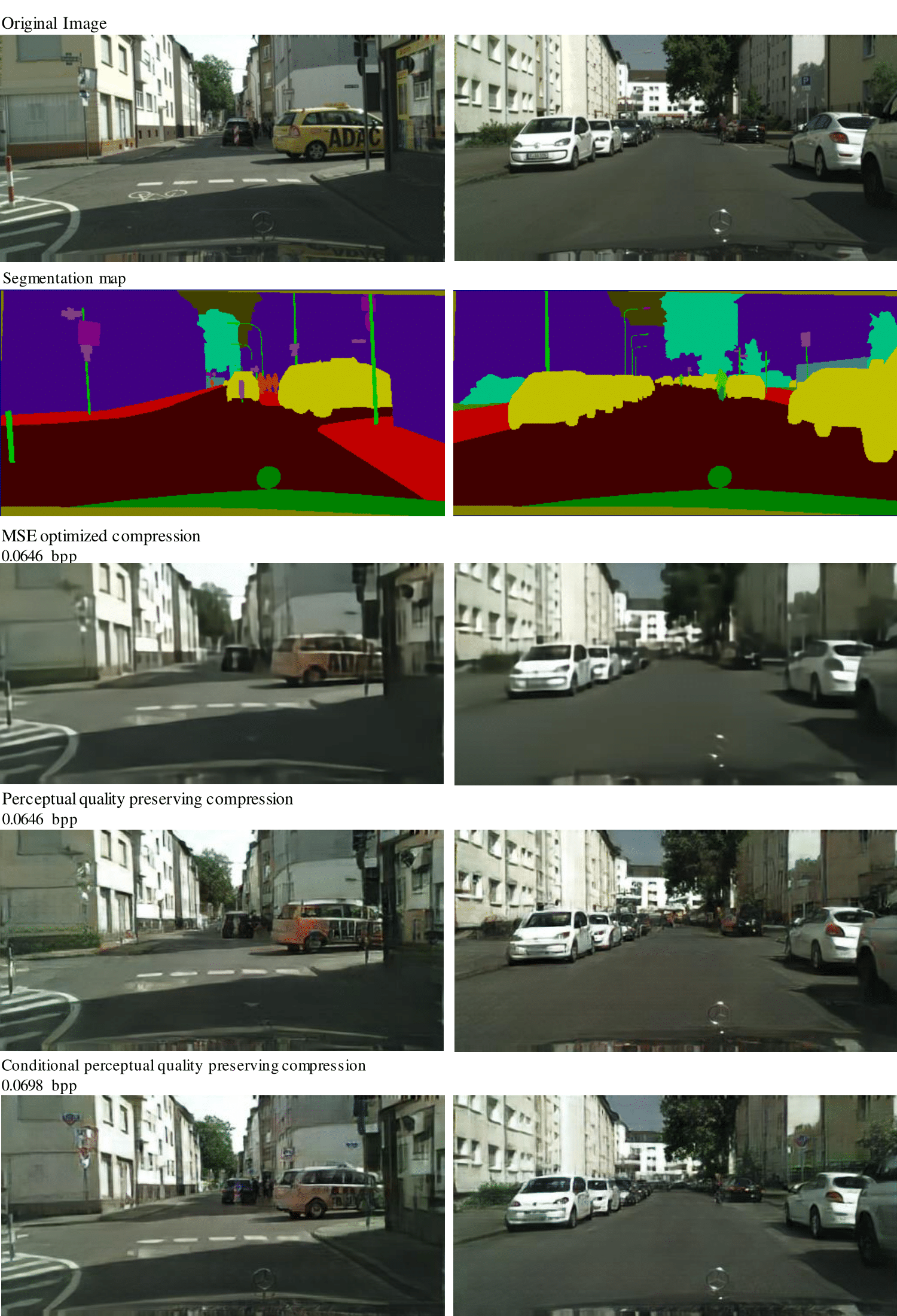}
 \caption{Qualitative results comparing original image, baseline MSE codec, perceptual quality preserving codec \citep{Yan2021OnPL} and our proposed conditional perceptual preserving codec for Cityscape dataset with rate $\approx 0.06$ bpp.}
 \label{fig:ctsp3}
\end{figure}

\begin{figure}[htb]
\centering
 \includegraphics[width=1.0\linewidth]{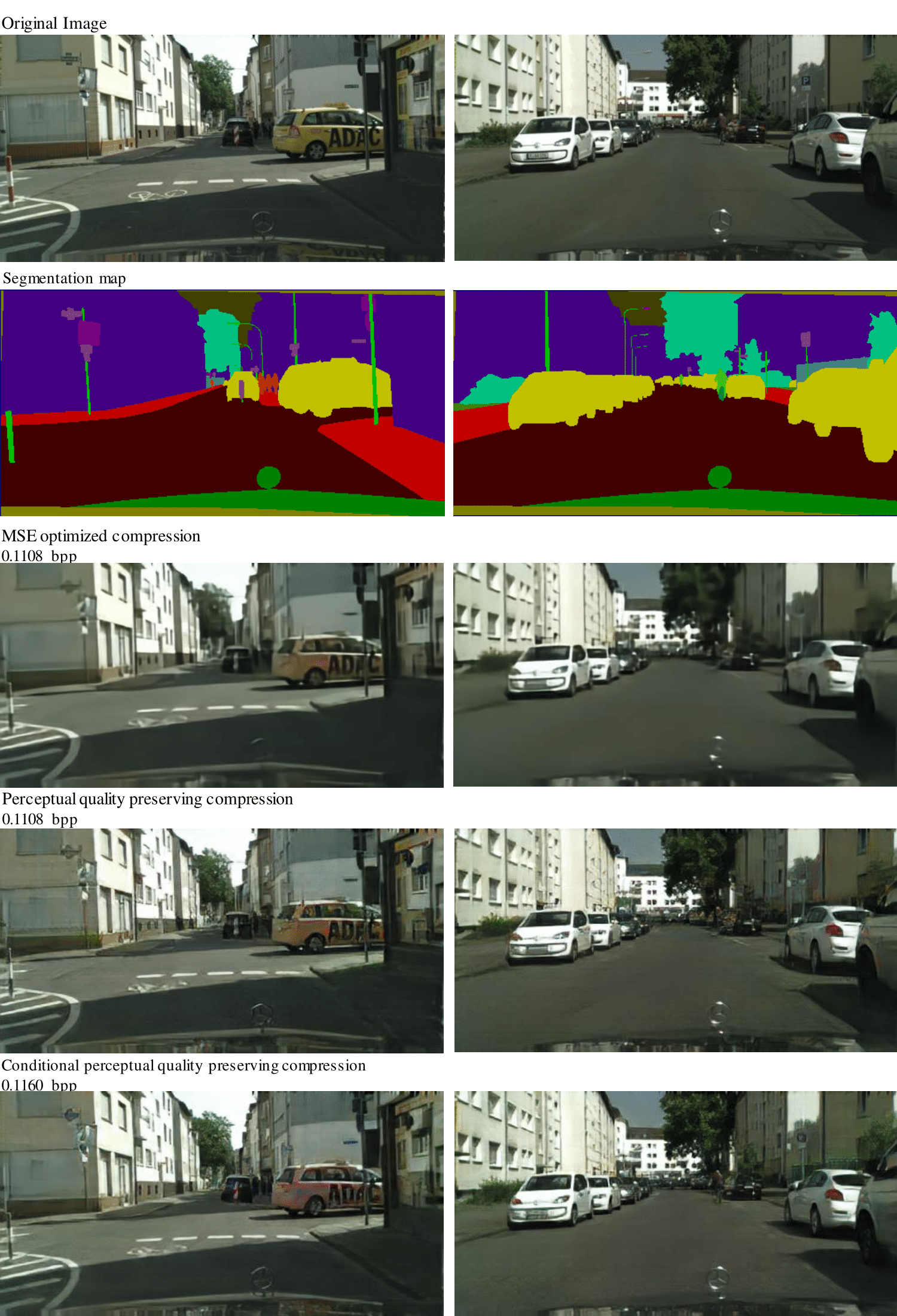}
 \caption{Qualitative results comparing original image, baseline MSE codec, perceptual quality preserving codec \citep{Yan2021OnPL} and our proposed conditional perceptual preserving codec for Cityscape dataset with rate $\approx 0.1$ bpp.}
 \label{fig:ctsp4}
\end{figure}

See more qualitative results for MNIST dataset in Fig.~\ref{fig:mnist1}-\ref{fig:mnist5}. See more qualitative results for Cityscape dataset in Fig.~\ref{fig:ctsp1}-Fig.~\ref{fig:ctsp4}.

\subsection{Evaluation on Common Randomness Required}
\label{app:evcr}
We verify the correctness of Theorem 4 by adjusting the amount of common randomness $H(W)$ and observe its impact on perceptual quality. The MNIST dataset is used and the experimental setup is the same as Section. 4.2. We constrain the noise $W$'s entropy that is injected in $g_2$ and observe its effect on FID. The $H(X)$ is approximated by the lossless compression rate of MNIST. We directly take the result from \citep{townsend2018practical} and assume that $H(X)\approx 1105.44$ bits.

More specifically, we use fully factorized unit Gaussian noise, which means that $W\sim \mathcal{N}(0,I)$. As the quantization of float32 is uneven, we can not directly obtain entropy $H(W)$ from the differentiate entropy of $\mathcal{N}(0,I)$. Therefore, we manually estimate the $H(W)$ by Monte Carlo. Specifically, we draw sample $w \sim \mathcal{N}(0,1)$, and find its previous float32 $w_{-1}$, its next float32 $w_{+1}$, and approximate its support as $\delta = (w_{+1}-w_{-1})/2$. Next, we estimate the probability mass of $w$ as $p(w) = cdf_{\mathcal{N}(0,1)}(w + \delta/2) - cdf_{\mathcal{N}(0,1)}(w - \delta/2)$, and the information as $-\log p(w)$. We repeat this process many times and obtain the average information. According to the weak law of large number \citep{cover1999elements}, this average converges to true entropy. We sample $10^{6}$ times, and obtain an average entropy of $26.55$ bits per dimension.

\begin{figure}[htb]
\centering
 \includegraphics[width=0.5\linewidth]{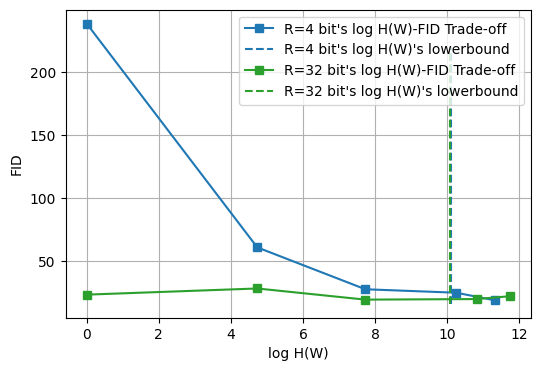}
 \caption{The effect of common randomness on perceptual quality. We define $\log H(W) = 0$ instead of $-\infty$ as deterministic.}
 \label{fig:crq}
\end{figure}

We gradually add noise $W$ to perceptual decoder $g_2$ from deterministic to $H(W) \ge H(X) - (R^{0}+1)$. The $R^0$ is the rate of codec, which is set to $4$ and $32$. As shown in Fig.~\ref{fig:crq}, for low bitrate such as $4$ bits, it is quite clear that an insufficient amount of noise entropy $H(W)$ severely affect the perceptual quality (FID $238.77$ vs $18.76$). However, for high bitrate, even deterministic codec can achieve a high perceptual quality (FID $23.23$). And the advantage of perceptual quality for sufficient noise beyond the lowerbound is not as obvious as low bitrate (FID $19.25$ vs $23.23$). Theoretically, it is necessary to have $H(W)\ge H(X) - (R^{0}+1)$ for perfect perceptual quality. Practically, we can not achieve perfect perceptual quality due to the capacity or optimization of generative model. Therefore, when bitrate is high, the effect of insufficient noise entropy $H(W)$ might not obvious.

\section{More Discussions}
\label{app:dis}
\subsection{Broader Impacts}
Saving bitrate without harming perceptual quality and semantic information in image has positive social impact. A large amount of energy and resources are spent on the transmission and storage of image data. Reducing the bitrate can save the resources, energy and the carbon emission during the process.

\subsection{Reproducibility Statement}
All theoretical results are proven in Appendix.~\ref{app:proof}. For experimental results, both MNIST and Cityscape dataset are publicly accessible. We provide implementation details in Appendix.~\ref{app:eset}. Furthermore, the source code for reproducing empirical results are provided in supplementary materials.

\end{document}